\documentclass[draft]{agujournal2019}
\usepackage{url} %this package should fix any errors with URLs in refs.
\usepackage{lineno}
\usepackage{soul}
\usepackage{amsmath,amssymb}
\usepackage[export]{adjustbox}

\draftfalse

\journalname{JAMES}

\begin{document}

%% ------------------------------------------------------------------------ %%
%  Title
%
% (A title should be specific, informative, and brief. Use
% abbreviations only if they are defined in the abstract. Titles that
% start with general keywords then specific terms are optimized in
% searches)
%
%% ------------------------------------------------------------------------ %%

% Example: \title{This is a test title}

\title{Quantifying Convective Aggregation Using the Tropical Moist Margin's Length}

%% ------------------------------------------------------------------------ %%
%
%  AUTHORS AND AFFILIATIONS
%
%% ------------------------------------------------------------------------ %%

% Authors are individuals who have significantly contributed to the
% research and preparation of the article. Group authors are allowed, if
% each author in the group is separately identified in an appendix.)

% List authors by first name or initial followed by last name and
% separated by commas. Use \affil{} to number affiliations, and
% \thanks{} for author notes.
% Additional author notes should be indicated with \thanks{} (for
% example, for current addresses).

% Example: \authors{A. B. Author\affil{1}\thanks{Current address, Antartica}, B. C. Author\affil{2,3}, and D. E.
% Author\affil{3,4}\thanks{Also funded by Monsanto.}}

\authors{Tom Beucler \affil{1,2,3}, David Leutwyler \affil{1,4}, Julia M. Windmiller \affil{1,4}}

\affiliation{1}{The authors contributed equally and their names appear in alphabetical order.}
\affiliation{2}{Department of Earth System Science, University of California, Irvine, CA, USA}
\affiliation{3}{Department of Earth and Environmental Engineering, Columbia University, New York, NY, USA}
\affiliation{4}{Max Planck Institute for Meteorology, Hamburg, Germany}

% \affiliation{4}{Fourth Affiliation}

%\affiliation{=number=}{=Affiliation Address=}
%(repeat as many times as is necessary)

%% Corresponding Author:
% Corresponding author mailing address and e-mail address:

% (include name and email addresses of the corresponding author.  More
% than one corresponding author is allowed in this LaTeX file and for
% publication; but only one corresponding author is allowed in our
% editorial system.)

% Example: \correspondingauthor{First and Last Name}{email@address.edu}

% \correspondingauthor{Tom Beucler}{tom.beucler@gmail.com}
% \correspondingauthor{David Leutwyler}{david.leutwyler@mpimet.mpg.de}
\correspondingauthor{Julia M. Windmiller}{julia.windmiller@mpimet.mpg.de}

%% Keypoints, final entry on title page.

%  List up to three key points (at least one is required)
%  Key Points summarize the main points and conclusions of the article
%  Each must be 140 characters or fewer with no special characters or punctuation and must be complete sentences

% Example:
% \begin{keypoints}
% \item	List up to three key points (at least one is required)
% \item	Key Points summarize the main points and conclusions of the article
% \item	Each must be 140 characters or fewer with no special characters or punctuation and must be complete sentences
% \end{keypoints}

% tgb - 1/22/2020 - Rewrote the key points to respect the 140 characters limit
\begin{keypoints}
\item The length of the margin separating moist and dry regions in the tropics can measure convective aggregation in simulations and observations.
\item We introduce a framework relating the moisture field's spatial organization to its time-evolution.
\item We infer an index quantifying convective aggregation in radiative-convective equilibrium simulations and the tropical Atlantic ITCZ.
\end{keypoints}

%% ------------------------------------------------------------------------ %%
%
%  ABSTRACT and PLAIN LANGUAGE SUMMARY
%
% A good Abstract will begin with a short description of the problem
% being addressed, briefly describe the new data or analyses, then
% briefly states the main conclusion(s) and how they are supported and
% uncertainties.

% The Plain Language Summary should be written for a broad audience,
% including journalists and the science-interested public, that will not have 
% a background in your field.
%
% A Plain Language Summary is required in GRL, JGR: Planets, JGR: Biogeosciences,
% JGR: Oceans, G-Cubed, Reviews of Geophysics, and JAMES.
% see http://sharingscience.agu.org/creating-plain-language-summary/)
%
%% ------------------------------------------------------------------------ %%

%% \begin{abstract} starts the second page

\begin{abstract}

On small scales, the tropical atmosphere tends to be either moist or very dry. This defines two states that, on large scales, are separated by a sharp margin, well-identified by the anti-mode of the bimodal tropical column water vapor distribution. Despite recent progress in understanding physical processes governing the spatio-temporal variability of tropical water vapor, the behavior of this margin remains elusive, and we lack a simple framework to understand the bimodality of tropical water vapor in observations. Motivated by the success of coarsening theory in explaining bimodal distributions, we leverage its methodology to relate the moisture field's spatial organization to its time-evolution. This results in a new diagnostic framework for the bimodality of tropical water vapor, from which we argue that the length of the margin separating moist from dry regions should evolve towards a minimum in equilibrium. As the spatial organization of moisture is closely related to the organization of tropical convection, we hereby introduce a new organization index (BLW) measuring the ratio of the margin's length to the circumference of a well-defined equilibrium shape. Using BLW, we assess the evolution of self-aggregation in idealized cloud-resolving simulations of radiative-convective equilibrium and contrast it to the time-evolution of the Atlantic Inter-Tropical Convergence Zone (ITCZ) in the ERA5 meteorological reanalysis product. We find that BLW successfully captures aspects of convective organization ignored by more traditional metrics, while offering a new perpective on the seasonal cycle of convective organization in the Atlantic ITCZ. 
% tgb - 2/25/2020 - Shortening the abstract to 250 words; commented out opening statement as we repeat it in the conclusion
% Overall, our new framework uses the tropical margin to connect the observationally-motivated, object-oriented view of convective organization to the spatio-temporal evolution of the water vapor field. It can be broadly deployed across models and observations, and offers an insightful and visual way to quantify convective organization. 

\end{abstract}

\section*{Plain Language Summary}
The tropical atmosphere tends to be either moist or very dry. This defines moist and dry regions that can be separated using a meandering line, namely the ``moist margin''. Better understanding the behavior of this ``moist margin'' would help explain the distribution of tropical water vapor, clouds and precipitation.
Here, we argue that the ``moist margin's'' length should evolve towards a minimum. By comparing the length of the ``moist margin'' to a theoretical minimum, we can assess how organized clouds and storms are in the atmosphere: The shorter the ``moist margin'', the more organized the clouds and storms. This simple rule allows us to better analyze the organization of idealized computer simulations of the tropical atmosphere as well as realistic datasets of the tropical Atlantic's atmosphere. 

\section{Introduction \label{sec:introduction}}

\begin{figure}
 \centering
    \noindent \includegraphics[width=\textwidth]{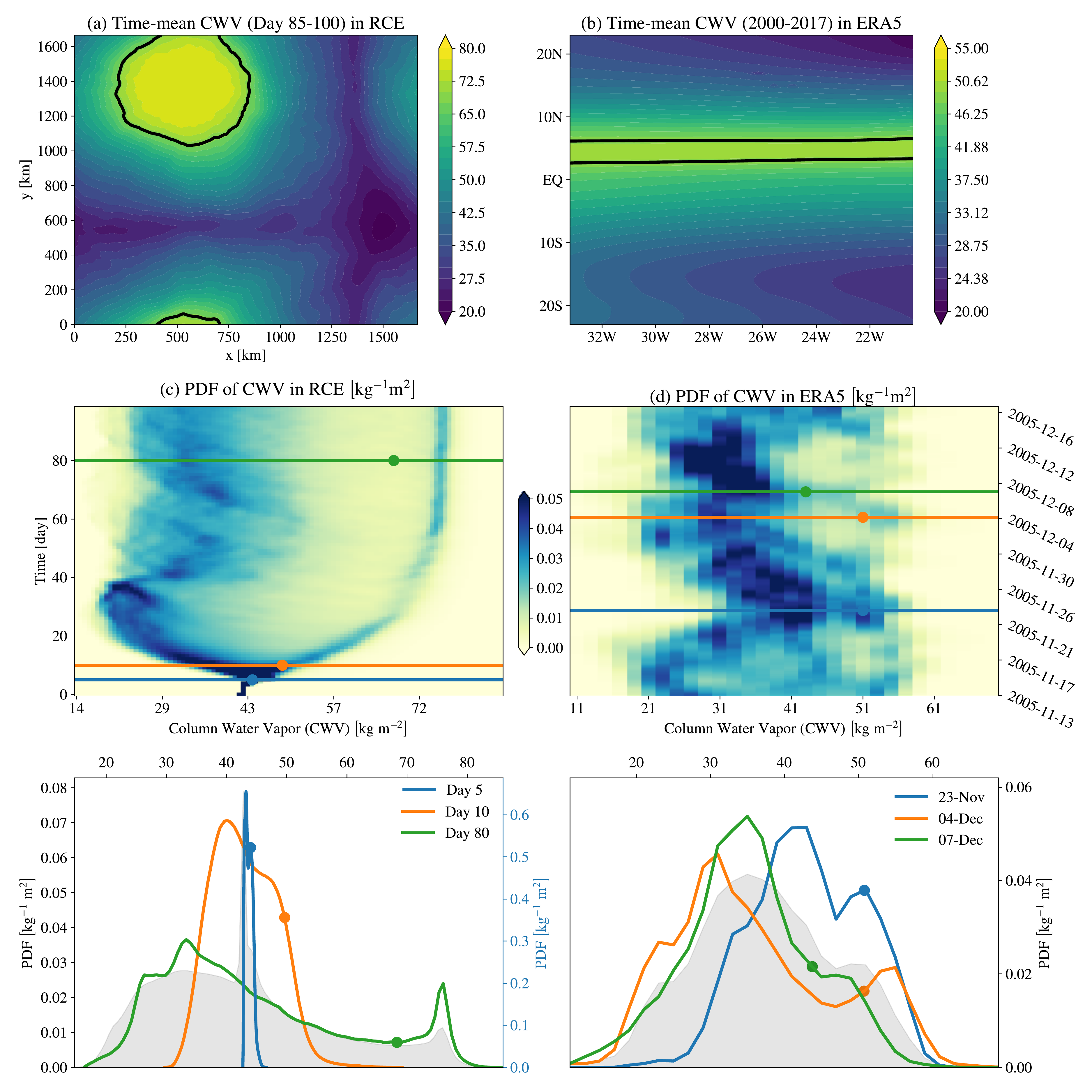}
 \caption{(Top row) Time-mean CWV in units $ \mathrm{kg\ m^{-2}}$ for (a) the last 15 days of the 300K RCE simulation and (b) the ERA5 reanalysis of the Atlantic Inter Tropical Convergence Zone (ITCZ) over the years 2000-2017. (Middle row) Time-evolution of the PDF of CWV for (c) RCE and (d) ERA5 (Nov-Dec 2005). For RCE, we denote the 88th percentile of CWV with dots, while we denote the 83rd percentile of CWV with dots for ERA5. (Bottom row) For RCE, we additionally sample the PDFs over 3 different 48-hour time-periods. For ERA5, we additionally sample the PDFs over 48-hour windows centered around the dates indicated in the top-right legend. In both cases, we add the time-mean PDF using gray shading. Note that we do \textit{not} claim that RCE is a good model for the ITCZ despite comparing both to show the versatility of our approach.}
 \label{fig:Fig1}
\end{figure}

The tropical hydroclimate is shaped by deep atmospheric convection, which occurs more frequently in the moist environments of other deep convective updrafts than in dry environments \cite<e.g.,>[]{Parsons2000, Redelsperger2002,peters2009mesoscale,sherwood1999convective}.
Deep convection may further moisten its environment through direct effects, such as increased detrainment of cloud water \cite<e.g., >[]{emanuel1996microphysical,sun1993distribution,minschwaner2004water,Holloway2009}, 
and indirect effects, such as vertical advection of water vapor driven by decreased radiative cooling in the presence of high clouds \cite<e.g.,>[]{bretherton2002simple,beucler2016moisture,wing2014physical}. 
In unforced simulations of radiative-convective equilibrium (RCE) initialized with a uniform humidity field, this positive feedback, referred to as convective self-aggregation \cite<e.g.,>[]{held1993radiative,tompkins1998radiative,bretherton2005energy,Wing_2017_review}, leads to the formation and maintenance of anomalously dry and moist regions (Figure~\ref{fig:Fig1}a and \ref{fig:Fig1}c). In the real atmosphere, bimodal probability density functions (PDF) of upper-tropospheric humidity have long been seen in satellite based observations of the mid-latitudes \cite<e.g.,>[]{soden1993upper,yang1994production} and the Tropics \cite<e.g.,>[]{zhang2003bimodality,brogniez2004humidite}. Since then, observational progress \cite<e.g., morphological compositing, see>[]{wimmers2011seamless} has confirmed the ubiquity of this bimodality for the full column water vapor (CWV) field (Figure~\ref{fig:Fig1}d).

Historically, satellite observations led to a \textit{Lagrangian} view of this bimodality, such as the advection-condensation framework  \cite<e.g.,>[]{sherwood1996maintenance,pierrehumbert1998evidence,sherwood2006distribution,beucler2016correlated} in which air masses are transported along isentropic surfaces between the Equator and the Poles. Their humidities correspond to the last time air masses reached saturation \cite<the ``cold trap'', e.g.>[]{pierrehumbert2007relative}, explaining the dry peak as air masses coming from the Poles and the moist peak as air masses coming from the deep Tropics. As the advection-condensation framework ambiguously mixes large-scale dynamics with microphysics, whose spatial scales are separated by $\sim $6 orders of magnitude, systematic analyses of idealized simulations were conducted to better isolate the respective effects of advection, convection, radiation, and microphysics on water vapor. This led to an \textit{Eulerian} view of this bimodality, in which the formation of large-scale moist and dry regions happens within a week to a month at the $\sim $1,000km-10,000km scale through the interaction between CWV and radiation in the slowly-rotating Tropics
%, where buoyancy gradients are weak 
\cite<e.g.,>[]{emanuel2014radiative,holloway2016sensitivity,beucler2019comparing,muller2015favors}. 
% tgb - 2/17/20 - Commented following Julia's comments on how this sentence may be confusing while not being critical to the introduction
%Mesoscale convective clusters form in O(1day) at the O(100km) scale when the vertical velocity and heating profiles are bottom-heavy, and the import of moist static energy (MSE) via low-level moisture convergence exceeds the export of MSE via mid-level temperature divergence \cite<e.g.,>[]{raymond2009mechanics,inoue2015gross,chikira2014eastward}. 

Recently, a \textit{mixed Eulerian-Lagrangian} perspective emerged. \citeA{Mapes2018} used blended satellite data to identify the 48-mm\footnote{$1\mathrm{mm}=1\mathrm{kg\ m^{-2}} $ when used to describe CWV} CWV isoline as a sharp margin, namely the tropical moist margin (e.g. the black solid line in Figure~\ref{fig:Fig1}b), separating two spatially-coherent regions: the ``dry tropics'' (dry maximum of the PDF) and the ``moist tropics'' (moist maximum of the PDF). The CWV distribution of a given air mass can then be inferred from the meandering dynamics of the moist margin, making this new Eulerian–Lagrangian framework \cite<see>[]{masunaga2020mechanism} suitable to both complex observations and idealized simulations \cite<see review by>[]{holloway2017observing}. These encouraging results, along with the tight link between CWV and convection in the Tropics, motivate the central question of our manuscript:

\textit{How can we leverage the time-evolution of the moist margin to better understand the organization of tropical CWV and convection?}

Here, we show that the tropical moist margin length (MML) can be used to quantify the degree of convective aggregation in both idealized RCE simulations and meteorological reanalysis. We introduce the datasets in Section \ref{sec:MethodsAndData}. We then generalize the coarsening framework of \citeA{craig2013coarsening} and \citeA{windmiller2019universality} to link the CWV's distribution to its time-evolution using the moist margin. Based on the coarsening framework, we argue in Section \ref{sec:Theory} that (i) we can diagnose the time-evolution of CWV using an empirical potential, an empirical Gibbs free energy, or simply the MML; and (ii) that the MML decreases the more aggregated CWV becomes. Afterwards, we derive a simple index exploiting the length of the moist margin to quantify convective aggregation in both idealized simulations and reanalysis in Section \ref{sec:object_oriented} before concluding in Section \ref{sec:Conclusion}.

\section{Data\label{sec:MethodsAndData}}

We analyze two high-resolution data sets with varying degrees of realism and different domain geometries: Idealized simulations of RCE on the one hand (Section \ref{sec:COSMO}), and high-resolution reanalysis of the tropical atmosphere on the other hand (Section \ref{sec:ERA5}). For conciseness, we do not introduce simulations of intermediate degree of realism, and contrast both data sets to illustrate the versatility of our framework rather than compare RCE to observations in details.

\subsection{Simulations of Non-Rotating RCE \label{sec:COSMO}}

We use the non-hydrostatic limited-area weather and climate model COSMO {{\cite<Consortium for Small-scale Modeling model, v5.0, >[]{Steppler_2003}}} in a square, doubly-periodic computational domain of 506$\times$506$\times$74 grid points with horizontal grid spacing of 3\,km, and use a time step of 30\,s. We conduct nine simulations with prescribed uniform SSTs ranging from 296\,K to 304\,K at 1\,K increments. Apart from the size of the computational domain, the configuration follows the RCE-MIP protocol \cite{Wing_2018}. As defined in the protocol,  each simulation is initialized with a typical tropical profile obtained from a small-domain RCE simulation, and run to stationarity (100 days). The reader interested in the details of the numerical model is referred to \ref{app:RCE_simulations}.

In three-dimensional, cloud-resolving simulations of non-rotating RCE, the CWV field evolves from an initially horizontally-homogeneous state into distinct, spatially-coherent, dry and moist regions (Figure~\ref{fig:Fig1}a) that map to a bi-modal distribution (Figure~\ref{fig:Fig1}c).
%\juliainline{evolution towards a bimodal distribution}
%\juliainline{include Figure 1c}
% tgb - There's no figure 1e/1f (see slack convo)
Qualitatively, the presented simulations evolve similarly to other unforced simulations of RCE  \cite<see, e.g.,>[ for a review]{Wing_2017_review}. 
A peculiar aspect of the present simulations is that they do not exhibit a dry bias \cite<e.g., end of Section 3.3 from>[]{holloway2017observing}. In contrast to typical limited-domain RCE simulations, the dry mode of the CWV PDF is centered around 32\,kg/m$^{2}$, which compares well to the CWV distribution observed in the tropical Atlantic (see Figure~\ref{fig:Fig1}d). Note that repeating the presented simulations on a computational domain using 206$\times$206$\times$74 grid points results in a drier domain-mean profile (not shown). However, compared to the CWV climatology in the Atlantic, the moist mode has much larger CWV values.

\subsection{Reanalysis of the Atlantic Intertropical Convergence Zone\label{sec:ERA5}}

As a step towards understanding convective aggregation in the observed atmosphere, we complement the idealized simulations of RCE with meteorological reanalysis of the Atlantic ITCZ.  

We choose the European Centre for Medium-Range Weather Forecasts Reanalysis
version 5 (ERA5), which was produced by assimilating
observational data in version CY41R2 of the Integrated Forecast System \cite{Hersbach2020}, for its high spatiotemporal resolution and realistic hydrologic cycle. This new reanalysis has a spatial grid of $0.25^{\circ}\times0.25^{\circ} $, which resolves some deep convective events and allows us to capture the moist margin's curvature. Furthermore, it has a temporal resolution of 1\,hour, allowing us to accurately sample the atmospheric energy budget over periods as short as 48 hours. 
The analysis period is defined as the first of January 2000 until the 31st of December 2017. 
To focus on the Atlantic ITCZ, we restrict our analysis to the domain between latitudes of 23S and 23N (the standard definition of the Tropics), and extending from longitudes of 34W to 14W. The domain, depicted in Figure~\ref{fig:Fig1}b, is located between the moistest regions of South America and Cape Verde, excluding land areas. 

The defined Atlantic domain is characterized by the very moist Intertropical convergence zone and by the much drier sub-tropics alike \cite{Mapes2018}. These two regions differ distinctly in their CWV, as  illustrated by the multiyear time-average of the CWV field, which exhibits a region of high CWV that zonally spans the domain (moist tropic), and a comparatively dry region surrounding it (sub-tropics), see Figure~\ref{fig:Fig1}b. The separation of the domain into dry and moist region is also evident in the CWV distribution, which is characterized by two modes (Figure~\ref{fig:Fig1}d). Its evolution displays alternating phases in which the PDF either becomes more uni-modal or more bi-modal. These different phases will be discussed in detail below.

\section{Theory\label{sec:Theory}}

%DL: Alternative proposal below:
%While the high-resolution of both datasets unambiguously resolve the bimodal structure of the CWV's distribution (Figure~\ref{fig:Fig1}c-d), it can be challenging to identify a single \textit{mechanism} explaining its time-evolution given the complex interplay between CWV, convection and precipitation.

Motivated by the results of self-aggregation studies, the key premise of this paper is that the degree of convective aggregation is reflected in the bimodality of the CWV distribution, with a more pronounced bimodality corresponding to a higher degree of aggregation.
Although three-dimensional simulations of RCE, satellite observations and reanalysis datasets display a bimodal distribution of CWV (Figure~\ref{fig:Fig1}c-d), a \textit{physical mechanism} unambiguously explaining this bimodal nature remains to be identified. In particular, testing hypotheses established using the RCE framework on observational datasets remains difficult as we lack analysis frameworks applicable to both. 
Thus, we here aim at developing a flexible yet simple framework relating the CWV's distribution to its time-evolution. 
Our theoretical framework is motivated by an analogy between the time-evolution of the tropical atmosphere's moist and dry regions and a phase separation process. We introduce this analogy (Section \ref{sub:analogy}), develop a flexible framework relating the CWV's distribution to its time-evolution (Sections \ref{sub:Potential} and \ref{sub:LFE}), and finally demonstrate that this relation can be described using a \textit{single variable}, namely the moist margin's length (MML) (Section \ref{subsec:Contour_length}). 

\subsection{A Phase Separation Analogy\label{sub:analogy}}

The analogy is that in the tropical atmosphere, moist and dry regions separate in a way similar to oil and water. Starting with a system that contains oil and water molecules, will the equilibrium state be a homogeneous mixture or two separate phases? At constant temperature and pressure, the equilibrium state minimizes the Gibbs free energy ($G$), which is the maximum amount of non-mechanical work that can be extracted from a closed system. In the case of an idealized mixture of oil and water, $G$ helps explain why the homogeneous mixture state is only favorable for temperatures far above the boiling temperature of water \cite<e.g.,>[p.~190]{Schroeder1999}.

According to the Landau theory of phase transitions, $G $ can be written as the sum of (1) a term only depending on state variables (e.g. temperature and pressure); and (2) a potential $V$ depending on a so-called ``order parameter''. In the case of an idealized oil and water mixture, the ``order parameter'' may be the oil concentration calculated within a finite mixture volume. For a fixed thermodynamic state, $G$ only depends on the order parameter, and this dependence can be entirely described using $V$: if the potential has two minima (\textit{double-well} potential), the equilibrium state will be two unmixed phases, while it will be a homogeneous mixture if the potential only has one minimum (\textit{single-well} potential). 

For a closed system, the second law of thermodynamics implies that $G$ monotonically decreases in time until the system reaches the state of minimal $G$, i.e. the equilibrium state. In the case of a double-well potential describing an oil and water mixture, $G$ monotonically decreases as oil and water separate. The order parameter's time-evolution is then given by the so-called time-dependent Ginzburg-Landau equation. In the case of a well-mixed but out-of-equilibrium system, this evolution is referred to as coarsening, which e.g. describes the formation and growth of isolated patches of oil and water in a salad dressing. One caveat, which will become important in our discussion of the ITCZ reanalysis data below, is that in the case of an open system, $G$ can also change via advection through the system's boundaries. Keeping this caveat in mind, we henceforth develop our theoretical framework using the closed RCE simulations before applying it to the ITCZ reanalysis.

\subsection{Potential\label{sub:Potential}}

Motivated by the well-resolved self-aggregation of convection and the corresponding broadening of the CWV distribution in our idealized RCE simulations, we start by asking the question: Why do anomalously moist and dry regions dry or moisten further? 

While several physical mechanisms supporting such a development have been formulated \cite{Tompkins2001c,bretherton2005energy,emanuel2014radiative,beucler2018linear,emanuel2019inferences}, \citeA{windmiller2019universality} show that these different mechanisms may all lead to similar upscale growth of moist and dry regions as long as humidity anomalies are locally reinforced. 
Whether local humidity anomalies are reinforced can be conveniently investigated by first making the following assumption:
%A local reinforcement of humidity anomalies can be conveniently described by making a key assumption: 
\textit{The time-evolution of CWV, described by its local time-derivative $ \partial_{t}\mathrm{CWV} $, only depends on $CWV$}. In other words, we study how fast values of CWV increase or decrease, assuming that for a given CWV the sign and rate only depend on the CWV value itself. In this case the CWV local time-derivative (CWV tendency for short) can be described by a function $f(\mathrm{CWV})$ that only depends on CWV: 
\begin{equation}
    \label{eq:feedback}
    \frac{\partial\mathrm{CWV}}{\partial t}=f(\mathrm{CWV})
\end{equation}
Following Section \ref{sub:analogy} and \citeA{windmiller2019universality}, we express $f(\mathrm{CWV})$ as the gradient of a (Landau) potential in CWV space, i.e.
\begin{equation}
    \label{eq:potential}
    \frac{\partial\mathrm{CWV}}{\partial t}=f(\mathrm{CWV})=-\frac{dV\left(\mathrm{CWV}\right)}{d\mathrm{CWV}}.
\end{equation}

% tgb - 8/2/2020 - Added a short reference above; between sections 3.1, 3.2, and 3.3., we already have a lot of qualitative text about this idea (compared to e.g. the ITCZ's time-evolution) and I would like to make sure the manuscript stays somewhat balanced
%Note, based on the idea of a Landau potential. 
%The upscale growth of CWV can be conveniently described by making a key assumption: \textit{The time-evolution of CWV, described by its local time-derivative $ \partial_{t}\mathrm{CWV} $, only depends on $CWV$}. In other words, we study which values of CWV change rapidly and which change slowly with time, and assume that the tendencies only depend on the CWV values themselves. Mathematically, this allows us to write the time-evolution of CWV as the gradient of a potential $ V\left(\mathrm{CWV}\right)$:
%\begin{equation}
%    \label{eq:potential}
%    \frac{\partial\mathrm{CWV}}{\partial t}=-\frac{dV\left(\mathrm{CWV}\right)}{d\mathrm{CWV}}
%\end{equation}
Under this assumption, the time-evolution of the CWV field is entirely determined by its potential, which describes the system's ability to stabilize by changing CWV. Provided with the potential $V(\mathrm{CWV})$, the question of whether local humidity anomalies are reinforced can then be addressed by investigating the shape of the potential. 

%\textbf{It is important to note that Eqs.~\ref{eq:feedback} and~\ref{eq:potential} should be considered as approximations, as in reality the tendencies of CWV may depend on more than just CWV itself.} 

There are, in principle, two ways to investigate the shape of the potential. 
On the one hand, we can start from physical principles and attempt to \textit{derive the potential} (or forcing function $f(\mathrm{CWV})$), as has been the focus of a number of previous studies \cite<e.g>[]{bretherton2005energy, craig2013coarsening, emanuel2014radiative}. 
On the other hand, we can directly use the data to \textit{diagnose the potential}, which is the approach we pursue in this work.  
To this end, we calculate an empirical potential for a given time-period and domain by first conditionally averaging all CWV tendencies (independently of time and location) based on their respective CWV content to obtain $f(\mathrm{CWV})$ and by then integrating equation~\ref{eq:potential} with respect to CWV:
%While previous work has attempted to derive a potential (or feedback function $f(\mathrm{CWV})$) from physical principles, we here use the CWV tendencies to diagnose an \textit{empirical} potential which describes the mean tendencies of CWV as a function of CWV.
%In particular, we can calculate an empirical potential for a given time-period and domain by first conditionally averaging all CWV tendencies (independently of time and location) based on their respective CWV content to obtain $f(\mathrm{CWV})$ and by then integrating equation~\ref{eq:potential} with respect to CWV:
% \textit{expected value} of the CWV tendency conditionally averaged on CWV:
\begin{equation}
V\left(\mathrm{CWV}\right)\approx-\int_{\mathrm{CWV_{0}}}^{\mathrm{CWV}}\mathbb{E}\left(\frac{\partial\mathrm{CWV}}{\partial t}|\widetilde{\mathrm{CWV}}\right)d\widetilde{\mathrm{CWV}},
\label{eq:calc_V}
\end{equation}
where $ \mathrm{CWV_{0}}$ is an arbitrary constant fixing the gauge of the potential so that $V\left(\mathrm{CWV}_{0}\right)=0$  and $ \widetilde{\mathrm{CWV}}$ is an integration variable. 
It is important to note that Equations~\ref{eq:feedback} and~\ref{eq:potential} are approximations, as in reality the tendencies of CWV may depend on more than just CWV itself. By assuming that the evolution of CWV is given by its mean value in CWV space, we successfully capture the time-evolution of the spatial-mean CWV field but reduce the spatial variance of the CWV time-tendencies by a factor $ \sim$100 (as estimated using \citeA{beucler2019comparing}'s framework, not shown). Finally, note that it is possible to choose moist static energy (MSE) instead of CWV as the order variable in equation \ref{eq:potential}: in that case, equation \ref{eq:calc_V} can be applied to the MSE tendencies of individual physical processes, which helps link our framework to the MSE budget framework extensively used by the self-aggregation community (see \ref{app:Decomposition} for more details).

\begin{figure}
 \centering
    \noindent \includegraphics[width=\textwidth]{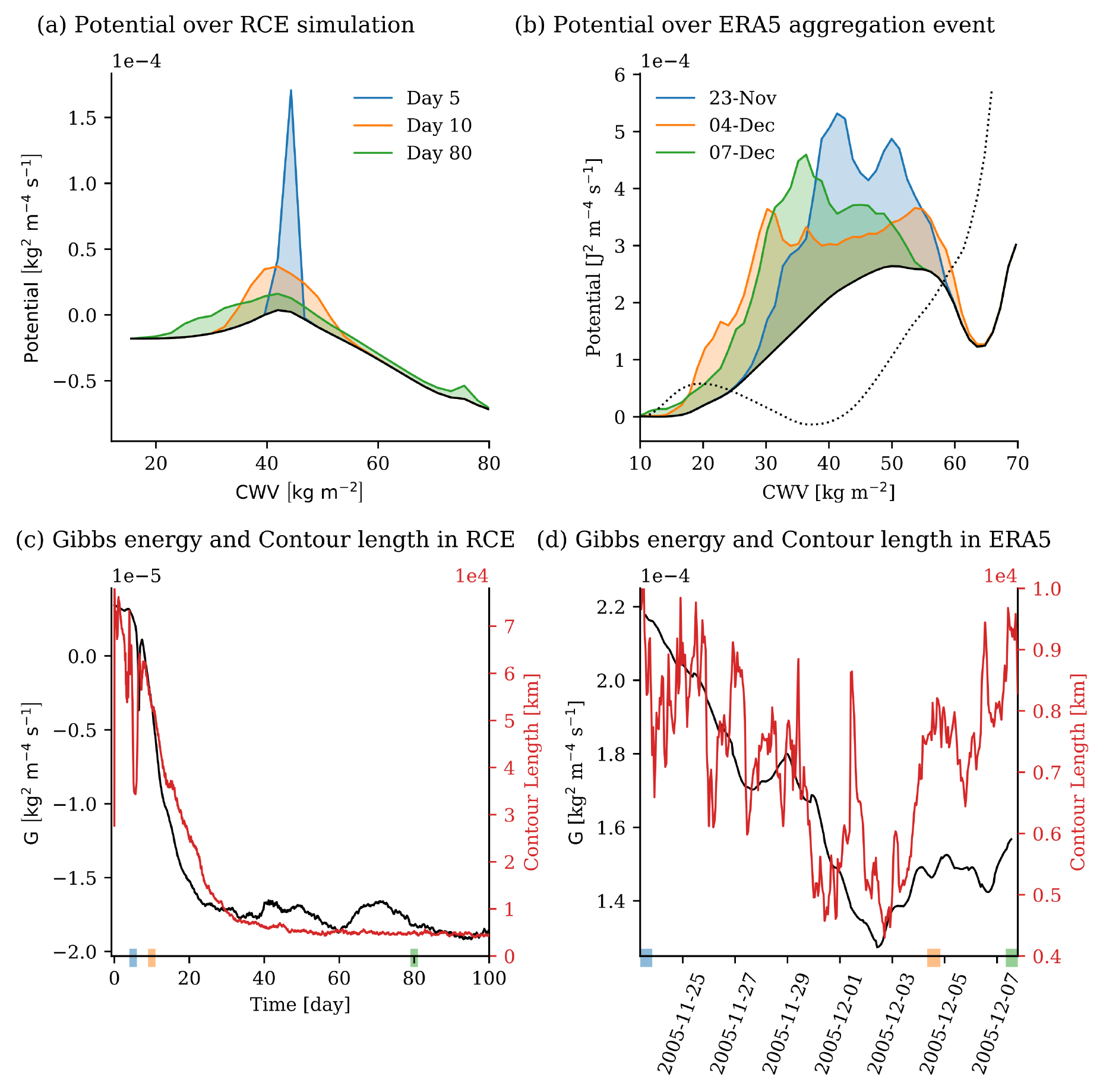}
 \caption{(a) Potential calculated from the CWV tendency in RCE (black line) and PDF of CWV sampled over three different time-periods (the shading indicates the sum of the potential and the PDFs, multiplied by a constant for visualization purposes). (b) Potentials calculated from the CWV tendency from ERA5 between 2005-11-23 and 2005-12-04 (solid black line) and between 2005-12-04 and 2005-12-07 (multiplied by a factor of 0.5, dotted black line) and PDF of CWV sampled over 48-hour windows centered around the dates indicated in the top-left legend (shading). (c) Empirical Gibbs free energy (black line) and contour length (red line) versus time for RCE. (d) Empirical Gibbs free energy (black line) and contour length (red line) versus time for ERA5. The time periods used in (a) and (b) are marked by the colored shading in (c) and (d).}
 \label{fig:Fig2}
\end{figure}

We are now in a position to use the two data sets described in Sec.~2 to illustrate how the shape of the potential can help us understand the time-evolution of the CWV distribution and therefore yield insight into the amplification of humidity perturbations in the tropical atmosphere. We start by discussing the potential and time-evolution of the CWV distribution for an idealized RCE simulation with a prescribed sea surface temperature of 300\,K (Figure~\ref{fig:Fig2}a).

We calculate the potential according to Equation~\ref{eq:calc_V} using the hourly tendencies of CWV over the 100 simulation days and 30 CWV bins. Its analysis reveals a double-well potential, i.e. a local maxium which separates two potential minima from each other. According to Equation~\ref{eq:potential}, the potential dictates the tendency for a given CWV to ``roll down'' towards the potential minimum and predicts faster moistening or drying for a steeper gradient. In the case of the presented double-well potential, the CWV content of atmospheric columns containing CWV values larger than the local maximum of the potential are predicted to increase, while the CWV content of atmospheric columns containing CWV smaller than the local maximum of the potential are predicted to decrease. CWV will continue to evolve until atmospheric columns reach the local minimum of the potential at the moist or dry ends of the CWV range. 
Note that it is difficult to determine the exact location of the two potential wells from Figure~\ref{fig:Fig2}a, which reflects challenges in sampling a potential minimum using Equation~\ref{eq:calc_V}. This sampling issue arises because the evolution of CWV towards the two minima becomes increasingly slow the closer the CWV gets to the potential minima as the gradient of the potential and hence the rate of moistening and drying go to zero (see Equation~\ref{eq:potential}). 
%(values smaller or larger than CWV$_{min}$ or larger than CWV$_{max}$ are, in fact, not expected at all). 
As a result, the two potential's local minima are calculated from a relatively small number of data points. However, despite the challenge of resolving the two potential's minima, the existence of a well-resolved potential maximum for a CWV value $\mathrm{CWV_{max}}$ indicates moistening of CWV values larger than $\mathrm{CWV_{max}}$ and drying of CWV values smaller than $\mathrm{CWV_{max}}$, as shown in Figures~\ref{fig:Fig2}a and b. Hence, a well-resolved potential maximum is sufficient to diagnose bistability as moistening and drying are physically limited by the positive-definition of CWV (lower bound of 0) and condensation of water vapor into liquid and ice (upper bound given by the saturation column water vapor). 

The evolution of CWV diagnosed by the potential compares very well with the time evolution of the CWV distribution shown in Figure~\ref{fig:Fig1}c.
This close link can be conveniently visualized by adding the CWV distributions shown in Figure~\ref{fig:Fig1}c to the potential. As the potential predicts that a given CWV content evolves either to the moist or dry potential minimum depending on its position relative to the potential maximum, the CWV distribution is expected to fill up the two potential wells with time. 
Starting from a CWV distribution which initially peaks at the potential maximum, we can see the two wells being filled up in Figure~\ref{fig:Fig2}a. 

Unlike in RCE simulations, CWV in the tropical Atlantic does not monotonically evolve towards a stationary state as it is continuously forced by CWV advection through all of the domain's boundaries. As a consequence, its CWV distribution can only be considered bimodal in the long-term time-average (see grey shading in Figure~\ref{fig:Fig1}d), while it frequently also appears to be uni-modal on $\sim $1week timescales (e.g., see Figure~\ref{fig:Fig1}d). This motivates us to split the CWV evolution into periods evolving towards a more bimodal CWV state (``aggregating'' phases), and periods evolving towards a more unimodal state (``dis-aggregating'' phases).

Repeating the analysis for a two-week period (from 2005-11-23 to 2005-12-07) in the tropical Atlantic shows a clear relationship between the ``aggregating'' phase of CWV and its potential (Figure~\ref{fig:Fig2}b). We have deliberately chosen this two-week period using the spatial variance of CWV so that it contains an aggregating phase (from Nov 23rd to Dec 4th) followed by a dis-aggregating phase (from Dec 4th to Dec 7th). In the first phase, the distribution widens considerably and appears to fill up the two potential wells, similar to the evolution observed in RCE. In the subsequent disaggregating phase, the system dries and the evolution does not follow the potential (drying of the moistest CWV and moistening of the driest CWV) anymore. 
The corresponding potential from the tendencies of the disaggregation phase (dotted black line in Figure~\ref{fig:Fig2}b) shows that, apart from a weak drying tendency for the extremely dry columns, all CWV values are predicted to evolve towards a single potential minimum around 40$\mathrm{kg\ m^{-2}} $. In contrast to the aggregating phase, the disaggregating phase is therefore more closely described by a single- rather than a double-well potential.

To summarize, using Equation~\ref{eq:calc_V} to calculate an empirical potential from the CWV tendencies provides a simple approach to assess the time-evolution of the CWV distribution. 
In particular, we have shown that the CWV distribution is expected to evolve towards a more bimodal distribution, with moistening of the moist and drying of the dry regions, if the associated potential has a double-well shape. 
%In particular, we have shown that a double-well potential structure diagnoses a moistening of moist and a drying of dry regions and thus an increasingly bimodal CWV distribution. 
Can we exploit these findings to define aggregating and dis-aggregating phases more formally?

% The potentials shown in Figure~\ref{fig:Fig2} are calculated from the full MSE tendencies and thus represent the combined effect of all processes which impact the MSE distribution. 
% By considering the impact from different processes on the MSE separately, we can also divide the overall potential into a potential for the separate processes to study their relative importance. 
% Using this approach, we find that the bimodality in the two potential shown in Figure~\ref{fig:Fig2} mainly results from XXX in the RCE simulation and from XXX (Figure~\ref{fig:Potential_decomposition}e) in the reanalysis XXX (Figure~\ref{fig:Potential_decomposition}f). 

% \tominline{Finish on the double well for smooth transition with the next sub-section?}

% \juliainline{Note that there is an interesting difference between the two potentials shown in Figure~\ref{fig:Fig2} regarding the depth of the potential wells. In particular, it is interesting to note that for the potential calculated for the RCE simulation the depth of the two potential wells is approximately the same while in the case of the reanalysis , the depth of the low MSE well is significantly deeper than the depth of the high MSE well. Studying the time evolution of the potential for a simple model, \todo{Craig_2013} note that the depths of the two potential wells become more and more comparable as the system approaches a steady state. The comparison of the two potentials in Figure~\ref{fig:Fig2} suggests, therefore, that the RCE simulations reach an almost steady state while the reanalysis, as might be expected, do not.}

\subsection{Gibbs Free Energy \label{sub:LFE}}

The double-well structure of the potential in idealized simulations of RCE suggests defining the degree of aggregation of a given CWV distribution as the degree to which the PDF of CWV has ``fallen'' into the two potential wells, which we calculate by taking the spatial-mean $\left\langle \cdot \right\rangle$  of the potential field $V\left[\mathrm{CWV}\left(t, \vec{x}\right)\right] $ over the spatial domain $\cal{D} $: 
\begin{equation}
  \label{eq:LFE}
  G\left(t\right)\overset{\mathrm{def}}{=}\left\langle V\left[\mathrm{CWV}\left(t, \vec{x}\right)\right]\right\rangle _{\cal{D}},
\end{equation} 
Expressed in words, $G$, at a given time, is the domain average of the empirical potential calculated from the CWV field. $G$ therefore quantifies the energy of a given configuration of the CWV field, and we refer to $G$ as the Gibbs free energy (or energy function) of our system in reference to the analogy in Section~\ref{sub:analogy}. Note that the the Landau-Free Energy defined by Equation 2 of \citeA{windmiller2019universality} is a special case of the Gibbs free energy defined here, as the Landau-Free Energy makes the additional assumption that all horizontal transport can be described by small-scale mixing, e.g. Fickian diffusion.

%It is interesting to note that, under two conditions, the Gibbs free energy defined in Equation~\ref{eq:LFE} corresponds to the Landau-Free Energy described by \citeA{windmiller2019universality} (their Equation 2). In particular, the LFE assumes that the potential acts only locally and that all transport is accomplished by diffusion. 
%Can be seen as the limit when the local amplification of humidity by processes like longwave radiation dominate and the transport can be modeled as random, small scale mixing. 

The time evolution of $G$ only depends on the time evolution of the CWV field. For instance, during an aggregating phase (i.e. during a phase in which the potential has a double-well shape), $G$ decreases as long as the CWV field becomes more bimodal. 
%Note that for a given potential $V$ in equation \ref{eq:LFE}, the LFE only depends on the CWV distribution. In other words, the LFE describes the evolution of the spatial average of the potential's local value, which is the average of the potential weighted by CWV.
It is interesting to note that if the time evolution of the CWV field were exactly described by Equation~\ref{eq:potential}, with a given potential function $ V\left(\mathrm{CWV}\right) $ fixed in time, $G$ would have to monotonically decrease in time. 
This can readily be shown by taking the time derivative of Equation~\ref{eq:LFE} \cite<see also e.g.,>[]{Krapivsky2010}:
\begin{equation}
    \frac{dG}{dt}=\left\langle \frac{dV}{d\mathrm{CWV}}\frac{\partial\mathrm{CWV}}{\partial t}\right\rangle_{\cal D}\overset{\text{\ensuremath{\left(\mathrm{Equation}2\right)}}}{=}-\left\langle \left(\frac{dV}{d\mathrm{CWV}}\right)^{2}\right\rangle_{\cal D}\leq0,
    \label{eq:minimize}
\end{equation}
where, in the first step, we have applied the chain rule and in the second step used Equation~\ref{eq:potential} to replace $\partial_{t} \mathrm{CWV} $ with the potential's gradient in CWV space. 
Note that Equation~\ref{eq:minimize} predicts opposite time-evolution's for the CWV distribution depending on whether the potential is bi-modal or uni-modal, i.e. whether we are in an aggregating or a dis-aggregating phase. In the case of a bimodal potential, $G$ is minimized by the moistening of moist and drying of dry regions while in the uni-modal case $G$ is minimized by the moistening of dry and drying of moist regions. 
%#L247
The monotonic decrease of $G$ predicted by Equation~\ref{eq:minimize} is, however, conditional on the time evolution of the CWV field being exactly captured by  Equation~\ref{eq:potential}. This in turn means that in a system where Equation~\ref{eq:potential} applies only approximately, as expected for the RCE simulations and the reanalysis data here, the time evolution of $G$ provides an  indication of how well a given potential describes the time evolution of the CWV field. In particular, periods during which $G$ increases indicate that the CWV field is not adequately described by Equation~\ref{eq:potential} during these periods.
Note, however, that in the case of an open system, e.g. in the reanalysis data, $G$ can also increase by advection through the boundaries (see Section~\ref{sub:analogy}).

The time evolution of $G$ for the RCE simulation and the reanalysis is shown in Figure~\ref{fig:Fig2}c and d, respectively. In the RCE simulation, $G$ rapidly decreases with time for about 40 days. After the initial rapid minimization, the decrease in $G$ slows down, until it ultimately starts oscillating around a new stationary state. As discussed in more detail below (Section~\ref{sec:object_oriented}), 
%#L258
the decrease in $G$ with time matches the increasing degree of aggregation in the RCE simulation. 
%this time evolution matches the increasing degree of aggregation with time in the RCE simulation. 
We also note that while the weak oscillation in $G$ during the later stages of self-aggregation cannot be explained within our framework (see Equation~\ref{eq:minimize}), they are reminiscent to oscillations previously seen in self-aggregation studies and discussed in detail by \citeA{Patrizio2019}.
In contrast to the RCE simulation, the time evolution of $G$ in ERA5 reanalysis is not as monotonic (Figure~\ref{fig:Fig2}b), but the two phases described above can still be clearly identified. In the first phase (aggregation phase, Nov 23rd to Dec 4th), $G$ decreases rapidly, and the CWV distribution becomes more bimodal. Note that the most bimodal state has a time lag of about two days with respect to $G$'s minimum. After reaching its minimum value on Dec 2nd, $G$ starts slowly increasing again, indicating that the CWV distribution is forced by a different potential that favors unimodality during the dis-aggregating phase (see e.g. Figure~\ref{fig:Fig1}b).

%To determine under what conditions we expect a coarsening like time-evolution from the above introduced empirical potential and Gibbs free energy, it is instructive to compare them with the potentials and the Landau-Free Energy discussed in \citeA{windmiller2019universality} (see their Eqs.~2 and~3). 
%Comparison shows that the key difference is the treatment of the horizontal transport of CWV. 
%While \citeA{windmiller2019universality} explicitly assume that all horizontal transport of CWV can be described by diffusion or small scale mixing and treat it separately from the purely locally acting potential, we make no assumption about the properties of how CWV is transported and include it into the empirically determined potential. 
%The condition under which we therefore expect coarsening is that the empirical potential decsribes local bistability and the horizontal transport of CWV can be mainly described as diffusive. 

It is important to note that while we use the time evolution of $G$ to diagnose when an equilibrium state is reached, we cannot use it to predict this equilibrium state's properties. 
According to Equation~\ref{eq:minimize}, a minimization of $G$ implies that the equilibrium state is only reached once the domain is either entirely moist or entirely dry, depending on whether the moist or dry potential well is deeper. However, this state might not be accessible because it violates energy or mass conservation. For example, isolated systems like doubly-periodic RCE simulations cannot have entirely dry (and therefore non-precipitating) equilibrium states as they would violate energy conservation: convective heating would be zero while radiative cooling would be non-zero. Additional constraints are required to make predictions about the final equilibrium state, such as the fractional area $\sigma $ of the domain in the moist and dry phases. We refer the interested reader to \citeA{lorenz2005convective,craig1996dimensional,craig2013coarsening} for simple $\sigma $ scalings from mass and energy conservation, and to \citeA{Renno_1996} for a historical $\sigma $ scaling that additionally models atmospheric convection as a heat engine.

In summary, our results suggest that for RCE simulations, the empirical Gibbs free energy $G$ can diagnose the time-evolution of the degree of aggregation quite well. For more realistic conditions, $G$ captures the aggregating and dis-aggregating phases and establishes a clear relation between $G$ and the modality of the CWV distribution. Although $G$ and CWV tendencies do not co-vary perfectly, the results are encouraging enough to simplify the analysis further. 

\subsection{Moist Margin's Length\label{subsec:Contour_length}}

%The concept of the Landau-Free Energy is well-rooted in statistical physics and successfully diagnoses the degree of aggregation as the CWV distribution becomes bi-modal. 
The empirical Gibbs free energy introduced above successfully diagnoses the degree of aggregation as the CWV distribution becomes bi-modal.
However, employing that diagnostic entails defining and calculating the full potential function, which requires $\sim $1week of hourly data. To further simplify our analysis, we aim to introduce a simple diagnostic variable that only requires a snapshot of the CWV distribution. 

For this purpose, we leverage the surface tension argument proposed in Section \ref{sub:analogy}, and argue that under the \textit{assumption of a strictly continuous CWV field}, $G$'s tendency is correlated with the length of the contour orthogonal to the gradient separating moist from dry regions, namely the moist margin (see Section~\ref{sec:introduction}). Analogous to the surface tension at the interface between water and air, which results from the larger attraction of water molecules to other water molecules rather than dry air molecules, the aggregation of moist regions may result from the increased likelihood of deep convection to occur in the moist environment of other deep convective updrafts rather than in the drier surroundings. The net effect of this attraction between deep convective updrafts is an inward force at the boundary of the moist tropics which acts to minimize the length of its contour, i.e. the moist margin's length (MML). 
The main point of this analogy is that a surface tension-like effect can result from the local reinforcement of humidity perturbations. 
Here, we investigate when, rather than why, humidity perturbations amplify and hypothesize that a surface tension-like effect acts during ``aggregating'' phases, i.e. during times of a double-well potential. 

To better apprehend the link between $G$ and MML, it is useful to understand how the atmosphere minimizes $G$ during aggregation: Initially, $G$ can be reduced by letting dry locations dry further (rolling down into the dry potential well), and moist locations moisten (rolling down into the moist potential well). A local drying and moistening would ultimately result in a spatially random field containing only dry and moist locations, each at the very bottom of the potential well. However, a strictly continuous field that spans across the two potential wells must always contain values in between, or it would not be continuous. Minimizing $G$, thus entails reducing the number of values on the potential hill as much as possible. In a continuous field, $G$ can thus only fully minimize after the local moist (or dry) patches merge. Combining the respective patches allows further reduction of the number of locations on the potential hill since fewer values are needed between the wells. Ultimately, the system should end up in a state with a single CWV minimum, a single CWV maximum, and a margin separating the two, orthogonal to a strong CWV gradient. Since the CWV value at the margin will always fall on the potential hill, estimating the length of the moist margin and its tendency from the CWV distribution should thus serve as an analogue diagnostic to $G$. 
In practice, the moist margin does not necessarily need to reside exactly on the peak of the potential hill, but likely it will reside "somewhere" on the hill. 

%\juliainline{highlight two points: potential maximum - antimode; height of antimode - contour length}

To estimate the MML from data, we proceed in 3 steps. First, we define the moist margin using a fixed percentile of the CWV distribution (e.g., the 88th CWV percentile in RCE, representing the antimode of the CWV distribution of the last 15 days; see \ref{app:percentile} for details). 
Choosing a fixed percentile allows the areas of the moist and dry regions to remain steady (assuming constant grid cell area) while the contour can freely evolve in time. 
This is advantageous for testing the idea of surface tension which, for a given area, predicts that the contour  length evolves towards the minimal possible length determined by the area. 
%Choosing a fixed percentile allows the areas of the moist and dry regions to remain steady (assuming constant grid cell area) while the contour can freely evolve in time, analogous to how volume is approximately conserved but surface area evolves in the case of the interface between water and air. 
Second, we use this moist margin to define a mask separating the dry regions (0) from the moist regions (1). Third, we measure the total contour separating zeros from ones using the measure module from the scikit-image library version 0.15.0 \cite{van2014scikit} in Python version 3.7.3 \cite<e.g.,>[]{sanner1999python}, subtract the contours at the domain's edges to avoid double counting of spatially-periodic regions, and multiply the result by the grid size to convert it to metric units. As a result, we can efficiently calculate MML at each timestep from a snapshot of the CWV field and the sole knowledge of the CWV percentile. Here, we choose the anti-mode of the bimodal CWV PDF as our CWV percentile, i.e. the local minimum between the two CWV maxima. Calculating this local minimum yields the 88th percentile for RCE and the 83rd percentile for ERA5.

We can now test the relationship between $G$ and MML by comparing their evolution in RCE simulations and ERA5 reanalysis (Figure~~\ref{fig:Fig2}c and d). For RCE, we find that the moist margin length systematically decreases in time and that there is a close (but not perfect) correlation between the time evolution of $G$ and the MML (correlation coefficient of 0.90). The correlation indicates that instead of $G$, we can also use the MML to measure the degree of aggregation. 
For ERA5 reanalysis, the time evolution of the MML is comparatively noisy, but it can still be divided into two periods. In particular, the MML shows an overall decrease during the aggregating phase in which $G$ decreases (correlation coefficient of 0.47), and an overall increase during the dis-aggregating phase in which $G$ increases (correlation coefficient of 0.64). These correlations of $\sim 0.5$ demonstrate that the link between $G$ and the MML is not as simple for externally-forced, open systems such as the real Tropics, where the time-evolution of $G$ and the MML cannot be related via a single surface tension constant. From a different perspective, the fast timescale of the MML's rapid oscillations and the slow timescale of self-aggregation are separated by a factor $\sim 20$ in RCE but only by a factor $\sim 10$ in ERA5 (see \ref{app:Timescales} for details), contributing to the weaker correlation between $G$ and the MML for ERA5. Nonetheless, the timescale separation and co-evolution of $G$ and the MML suggest that we can measure the degree of aggregation by measuring the MML, motivating the development of a new index for convective aggregation in the next section.

\section{Application to Idealized Simulations and Reanalysis\label{sec:object_oriented}}

%\juliainline{We will now use the moist margin length to investigate the degree of aggregation in the RCE simulation and the reanalysis. Based on the surface minimization idea discussed above, the most aggregated state is reached when the surface is minimized. Introduce index which compares the moist margin length to the minimal possible moist margin length: Def. of BLW In the following we will first use BLW to discuss the time-evolution of the moist and dry regions in the RCE simulations; additional phase of aggregation which mainly impacts the shape and number of moist regions which is not picked up by IQR. Second, we will use BLW to show that there the ITCZ exhibits a seasonal cycle of aggregation. }

So far, we have hypothesized that processes exist which make the tropical atmosphere bi-stable with respect to its humidity content and thus act to establish the bi-modal nature of column water vapor (CWV) in the tropics. 
In particular, we have outlined how this bi-stable nature can be understood in terms of a double-well potential, and how the replenishment of the potential wells relates to changes in the corresponding energy of the system ($G$). The observation that the tendency of $G$ correlates well with the tendency of the length of the contour separating dry and moist areas (MML) has led us to propose the moist margin length as a diagnostic for convective aggregation. Below, we illustrate how these ideas can be used to relate changes in the CWV distribution to changes in contour length. To this end, we make use of our two data sets: three-dimensional RCE simulations (Section \ref{sec:results_simulations}) and the Atlantic ITCZ (Section \ref{sec:results_ITCZ}). The latter is more complex as the RCE simulation depicts an idealized representation of the tropics using homogeneous boundary conditions and no planetary rotation on a square domain. This difference between the two systems is addressed by developing two distinct yet consistent definitions of the minimum contour length (Section \ref{sec:BLW_definition}).

% tgb - 8/1/2020 - This makes the transition less obvious; trying to move it to 4.3
% Furthermore, the double-periodic domain of the RCE simulation yields a horizontally closed system with no mass advection through the domain boundaries. In contrast, the Atlantic domain as represented in the re-analysis entails a horizontal mass flux thorugh the boundaries. 

\subsection{Definition of the BLW index   \label{sec:BLW_definition}}

Based on the ideas outlined in Sections \ref{sub:LFE} and \ref{subsec:Contour_length}, we introduce a new index to measure the degree of aggregation of a given CWV distribution. The key idea of the index is that aggregation leads to a shortening of the Moist Margin Length (MML), i.e., the total length of the contours surrounding the moist regions. 
% tgb - 2/6/2020 - The sentence below is vague and the assumptions come a bit later in the text
% Defining an index allows us to relate the MML to a reference state, but also involves system-specific assumptions.

\begin{figure}
  \centering
  \noindent\includegraphics[width=\textwidth]{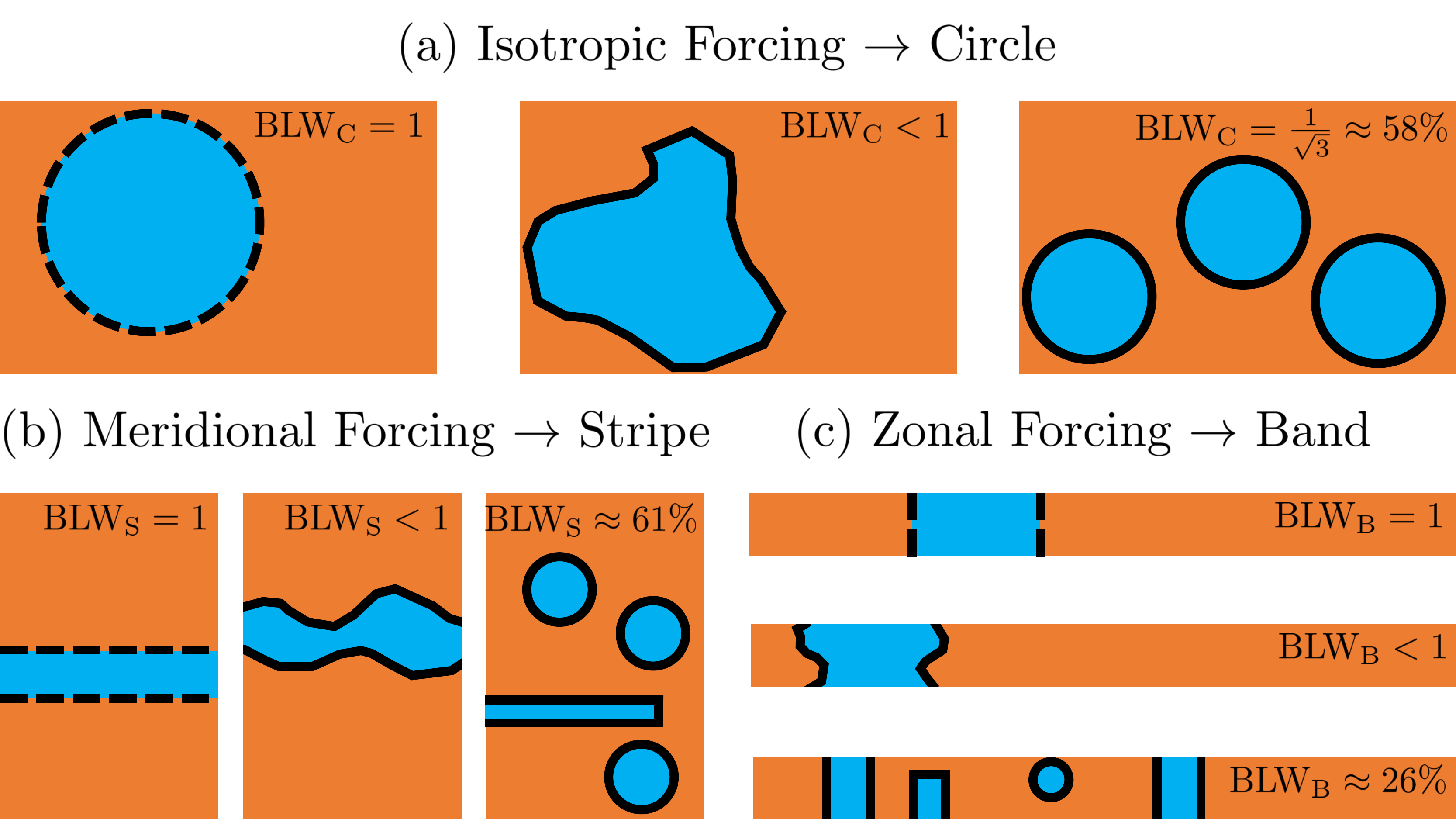}
  \caption{BLW for idealized convective clusters (light blue shapes) in a drier environment (orange background) for three different moisture forcing scenarios leading to three distinct shapes of minimal contour (dashed black lines): (a) Isotropic forcing leading to a circle, (b) Meridional forcing leading to a zonally-oriented stripe, and (c) Zonal forcing leading to a meridionally-oriented band.} 
  \label{fig:BLW}
\end{figure}

In general, the index is defined as:
\begin{equation}
    \mathrm{BLW}\overset{\mathrm{def}}{=}\left(\frac{\mathrm{Minimal\ Contour}}{\mathrm{MML}}\right)
\end{equation}
where we compare the Moist Margin Length (MML) to the Minimal Contour, i.e. the minimal length the contour could have for an assumed underlying shape of the moisture field. As illustrated in Figure \ref{fig:BLW}, an index $\mathrm{BLW}=1\ $indicates that the moisture field adopts the exact underlying shape of minimal contour. In contrast, $\mathrm{BLW}=0\ $ corresponds to the limit of an infinitely-long moist margin, either because there is an infinitely large number of small moist clusters or because the contour is non-rectifiable, i.e. ``fractal-like''. Intermediate values of BLW correspond to a ``deformed'' contour or a finite number of moist clusters.

For doubly-periodic RCE simulations on square computational domains, the minimal contour depends on the fraction of the domain covered by the moist region. As outlined by \citeA{holloway2016sensitivity}, if the moist region covers less than about one third (more precisely $100/\pi$ percent) of the domain, the minimal contour MML is obtained when the moist region is organized in the shape of a single circle of radius $ \mathrm{Minimal\ Contour}/(2\pi) $ and area $\cal A$, yielding the ``circle'' version of our index (Figure~\ref{fig:BLW}a): 

%For the double-periodic RCE simulations on a square computational domain, the minimal contour MML is obtained when the moist region is a single circle of radius $ \mathrm{Minimal\ Contour}/(2\pi) $ and area $\cal A$, yielding the ``circle'' version of our index: 

\begin{equation}
    \mathrm{BLW}_\mathrm{C}\overset{\mathrm{def}}{=}\frac{2\sqrt{\pi\times{\cal A}}}{\mathrm{MML}}. 
    \label{eq:BLW_C}
\end{equation}

The moist area (${\cal A}$) is defined as the total area of the grid points exceeding the CWV threshold used to define the moist margin. To define the moist margin, we use a temporally fixed percentile of the CWV PDF: the 83rd percentile for ERA5 and the 88th percentile for RCE (see \ref{app:percentile} for details). A notable advantage of using a percentile-based approach instead of an absolute threshold such as 48\,kg/m$^2$ is that it maintains a constant area for the wet and dry regions during the simulation's evolution and across the nine prescribed SSTs. In other words, the rank in the CWV distribution (and thus the area of the moist region) is fixed, while the threshold magnitude in units $ \mathrm{kg\ m^{-2}}$ changes as the simulation evolves, allowing us to meaningfully compare the degree of aggregation across SSTs.   

In contrast to the RCE simulations discussed above, the presence of a meridional SST gradient in the tropics externally organizes deep convection into the tropical rain belts. The predominantly zonal orientation of the tropical rain belt in the Atlantic is reflected in the zonal orientation of the moist region (see e.g. Figure~\ref{fig:Fig1}b).  In that case, the minimal contour MML is obtained when the moist region is a single zonal stripe of length MML/2, yielding the ``stripe'' version of our index (Figure~\ref{fig:BLW}b):  

\begin{equation}
    \mathrm{BLW}_\mathrm{S}\overset{\mathrm{def}}{=}\frac{2\times{\mathrm L}}{\mathrm{MML}}, 
\end{equation}

where L is the length of the considered domain. In the particular case of the Atlantic stripe, this length is approximately 2200km, where we have approximated the $ 1^{\circ}\times1^{\circ}$ longitude-latitude grid from ERA5 as a locally Cartesian $110\mathrm{km}\times110\mathrm{km} $ grid.

In the common case where wind shear organizes convection into ``bands'' orthogonal to the wind shear vector, namely squall lines and arcs \cite<e.g.,>[]{rotunno1988theory,robe2001effect}, the ``stripe'' version of our index can be generalized to take into account the forcing's orientation. For example, long-channel simulations of RCE typically develop strong zonal wind shears \cite<e.g.,>[]{posselt2012changes,wing2016self} that organize moist regions into meridional bands, motivating the ``band'' version of our index (Figure~\ref{fig:BLW}c):

\begin{equation}
    \mathrm{BLW}_\mathrm{B}\overset{\mathrm{def}}{=}\frac{2\times{\mathrm \ell}}{\mathrm{MML}}, 
\end{equation}

where $\ell $ is the width of the considered domain. 

Before applying BLW to our two datasets, we remind the reader that BLW only relies on two explicit assumptions in \textit{all} cases: (i) the CWV threshold used to define the moist margin; and (ii) the underlying moisture shape of minimal contour. These two assumptions make BLW values easy to interpret (Figure~\ref{fig:BLW}), especially when choosing the ``default'' values of the two assumptions: (i) the CWV PDF maximum if the PDF is unimodal and its anti-mode if the PDF is bimodal; and (ii) a circle if the forcing is isotropic or its direction unknown. For comprehensiveness, we finish this section by noting two corner cases that require adapting assumption (ii) and hence the definition of BLW. In the isotropic case, if the moist area occupies most of the domain, then the reference circle shape may be more appropriate for the dry than the moist region, in which case $\cal{A} $ in equation \ref{eq:BLW_C} must be replaced with the total area of the grid points \textit{below} the CWV threshold used to define the moist margin. In the anisotropic case, if there are not enough grid points exceeding the CWV threshold to form a moist stripe extending across the domain, then it might be more appropriate to assume a different reference moisture shape, e.g. a circle.
%\juliainline{I think the default idea of a circle should be mentioned here.}

\subsection{Idealized Simulations of Radiative-convective Equilibrium \label{sec:results_simulations}}

\begin{figure}
  \centering
  \noindent\includegraphics[width=\textwidth]{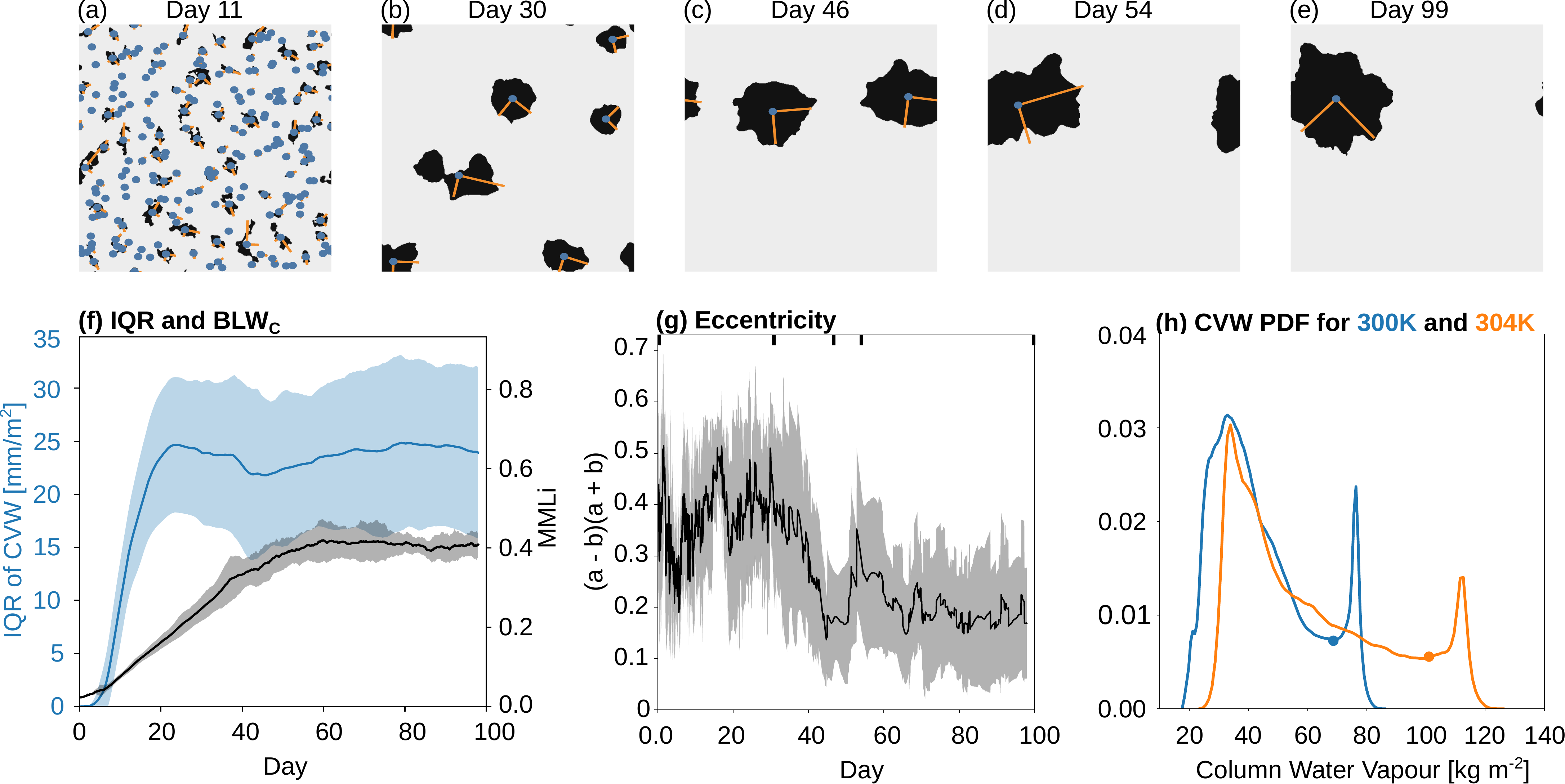}
  \caption{Evolution of CWV and aggregation diagnostics in a set of nine RCE simulations with varying SSTs (296\,K - 304\,K, depicted by the shading in f-g), after applying a 48\,h temporal mean. (a-e) Snapshots of high and low CWV in the 300\,K simulation: The black mask identifies grid points above the 88th percentile of CWV. The blue dots mark the centroids of connected grid points, and the orange lines the respective principle axis. (f) BLW$_\mathrm{C}$ (black) and CWV inter-quartile range (blue).  (g) Eccentricity of the largest object (mean in black, ensemble standard deviation in gray). (h) PDF of CWV for SSTs of 300 and 304\,K SST, averaged from day 85 to 100. The dots denote the 88th percentile.} 
  \label{fig:idealized}
\end{figure}

In the first few days of the simulation, the distribution of CWV is strongly peaked on top of the local potential maximum (see blue line in Figure~\ref{fig:Fig2}a). As soon as the first convective cells form,  self-aggregation starts acting on the moisture field, (negative $G$ tendency accelerates after day~11, Figure~\ref{fig:Fig2}c)  and the CWV distribution quickly becomes bi-modal (Figure~\ref{fig:Fig1}c). During this phase, a dry maximum can be identified around 32\,kg/m$^2$ and a moist maximum around 55\,kg/m$^2$. In the subsequent phase, the dry (or first) maximum remains at a constant value, while the moist (or second) maximum moves to beyond 70\,kg/m$^2$. Qualitatively, the development into a bi-modal distribution appears robust across the nine simulated SSTs (see \ref{app:percentile}). However, at warmer SSTs, the maxima of the two modes are further spaced apart. For example, the maxima are located at 33.2\,kg/m$^2$ and 75.9\,kg/m$^2$ at 300\,K, and at 33.9\,kg/m$^2$ and 111.9\,kg/m$^2$ at 304\,K (Figure~\ref{fig:idealized}h). 

The ratio of the dry to moist mode appears constant, despite the moist mode rapidly moving to larger CWV values and becoming less peaked with increasing SSTs. More specifically, the minimum separating the two modes remains at the 88th percentile of the CWV distribution (see dots in Figure~\ref{fig:idealized}h), i.e., the moist mode constantly covers about 1/9th of the computation domain. Here, we exploit this property to empirically define a contour dividing the domain into a dry and a wet area (see Section \ref{sec:BLW_definition}). 

Obtaining that contour allows for a spatial, object-oriented perspective of the simulation's evolution (Figure~\ref{fig:idealized}a-e): After the initial set of convective cells have formed, many small cells of high moisture content quickly coalesce into increasingly larger clusters. As they coalesce (days 11--30), the eccentricity of the largest object steadily increases, indicating that the clusters become increasingly asymmetric (until Day 30, see Figure~\ref{fig:idealized}g). Once only a few clusters remain, the eccentricity starts to decrease (i.e., the clusters become more circular), until they merge into a single cluster around day 54. 
Note that the merging process of the two last clusters leads to the transient formation of a very elongated cluster which can be seen as a sudden increase in eccentricity around day 54. During the next ten days, the final cluster becomes more circular until its shape becomes stable around day 65.

The described evolution appears well-captured by BLW$_\mathrm{C}$ (Figure~\ref{fig:idealized}f).  Between days 11 and 46, the overall contour length is rapidly shortened by the small moist cells coalescing and thus the index increases rather quickly. During the middle period (day 46--65), the rate of change in BLW$_\mathrm{C}$ slows down, as the contour is then mainly shortened by spatial re-arrangement of moisture towards a circular shape. In the final equilibrium phase, the value of BLW$_\mathrm{C}$ remains rather constant.
%\juliainline{I would move the discussion of IQR here (rather than having it in the conclusion)?}

Although the cluster at day 99 (Figure~\ref{fig:idealized}e), appears quite round, BLW$_\mathrm{C}$ only achieves a value of about 0.4, indicating that, at the considered time scales, the obtained contour remains more than twice as long as that of a perfect circle with the same area. Applying a 20-day running mean results in a CWV field similar to that displayed in Figure~\ref{fig:Fig1}a, and a BLW$_\mathrm{C}$ value of about 0.9. In the presented analysis, we applied a two-day running mean to the CWV field to filter short-lived convective cells.

Finally, it is interesting to compare the time evolution of BLW with the time-evolution of the Inter-quartile Range of CWV (IQR; 75th - 25th percentile) -- a widely-used, PDF-based index for assessing self-aggregation in simulations \cite<e.g.,>[]{bretherton2005energy,muller2012detailed,arnold2015global,holloway2016sensitivity}.
As increasing values of IQR indicate an increase in the degree of aggregation, the time evolution of IQR (Figure~\ref{fig:idealized}f) alone would suggest that the simulation already reaches its fully aggregated state after about 20 days. IQR can be shown to mainly capture the evolution of the dry mode, hence missing essential parts of the aggregation process (see Figure~\ref{fig:idealized}a-e), which are well-captured by BLW$\mathrm{_C} $. We find similar shortcomings when contrasting how BLW and the IQR of column relative humidity change with SST in \ref{app:percentile}.

% Draft IQR discussion
% That sensitivity of PDF-based indices becomes apparent when comparing BLW to the Inter-quartile Range (IQR, 25th - 75th percentile)  -- a popular index for assessing self-aggregation in simulations. In contrast to BLW, the IQR develops very rapidly, as it mainly grows with the development of the dry mode (Figure~\ref{fig:Fig1}c). Note that since the minimum separating the two PDF modes is represented by the 88th percentile, the 75th percentile does, by definition, not capture the moist mode. By blending a PDF-based method (considering the minimum separating the two modes) and an object-oriented method (contour length), the proposed index aims at overcoming these issues. 

%\juliainline{Not urgent but perhaps we want to replace \ref{fig:Fig1}a by the 20 day mean.}

\subsection{Reanalysis of the Atlantic Inter-tropical Convergence Zone\label{sec:results_ITCZ}}

In its stripe version (BLW$_\mathrm{S}$), the BLW index provides a tool to asses the evolution of the CWV field in the tropical Atlantic while yielding values comparable to BLW$_\mathrm{C}$ used for the RCE simulations. Below we illustrate these capabilities by extracting the seasonal cycle of aggregation of the Atlantic ITCZ using the 83rd percentile of CWV to define the moist margin. 

%Using 18 years (2000-2017) of ERA5 data, we calculate a BLW$_\mathrm{S}$ time-series from the hourly CWV fields for the stripe-like region over the tropical Atlantic, defined in Section~\ref{sec:ERA5}. 
%The seasonal cycle of BLW$_\mathrm{S}$ is calculated from the resulting time-series by binning the data using biweekly intervals, see Figure\ref{fig:seasonal_cycle}. 

\begin{figure}
    \centering
    \noindent \includegraphics[width=0.95\textwidth]{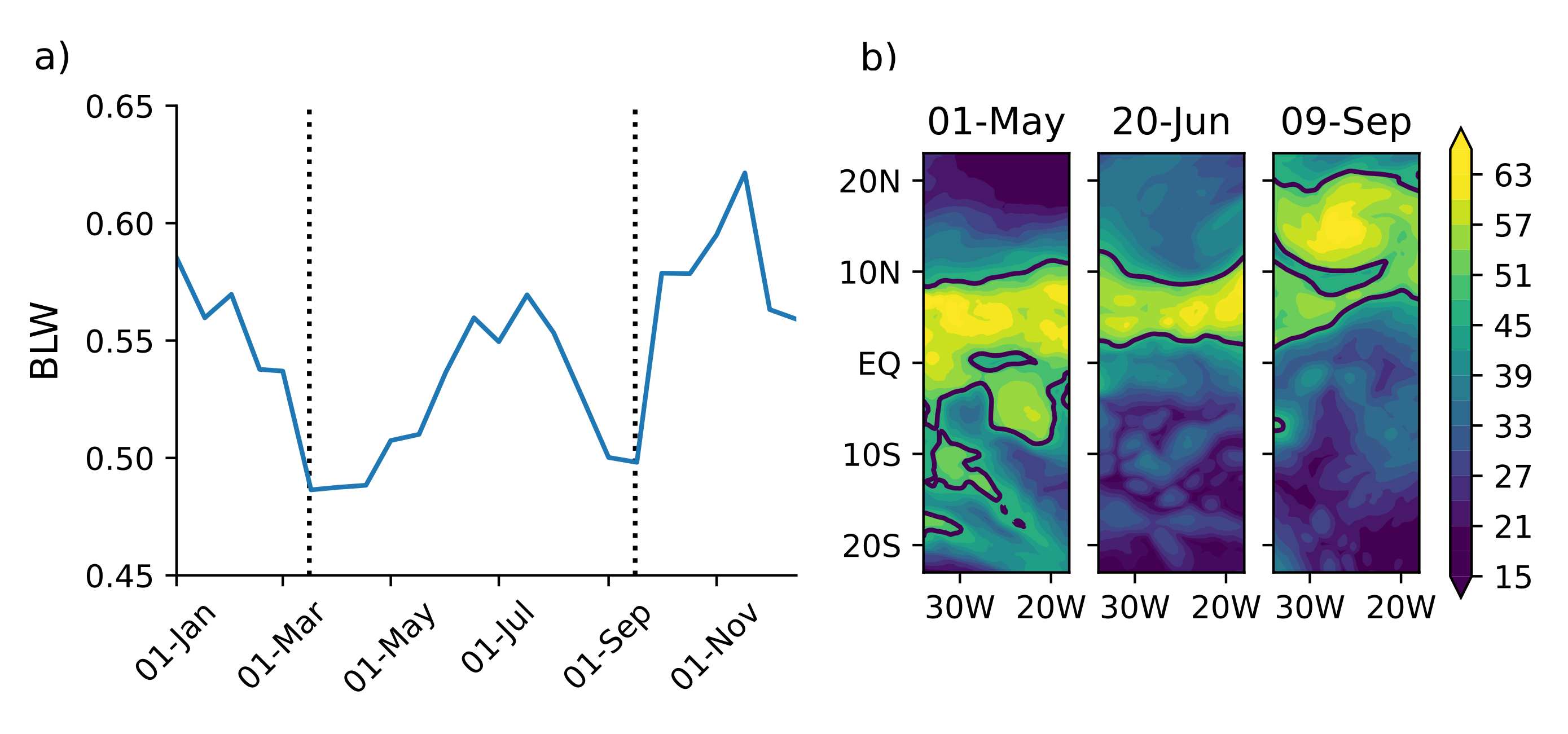}
    \caption{(a)  The annual cycle of $\mathrm{BLW_{S}}$ diagnosed for the 2000-2017 period. The two dotted lines mark March 16th and September 16th. (b) Snapshots of CWV for three days in 2010, representing the Atlantic ITCZ's annual cycle.}
    \label{fig:seasonal_cycle}
\end{figure}

The seasonal cycle of BLW$_\mathrm{S}$ exhibits two minima indicating two periods during which the Atlantic ITCZ is in a comparatively dis-aggregated state, one during boreal spring (local minimum on March 16th) and one during boreal fall (local minimum on September 16th), see Figure~\ref{fig:seasonal_cycle}. To obtain the BLW$_\mathrm{S}$ climatology, the index has been computed on the hourly CWV snapshots of the ERA5 reanalysis, and then binned into biweekly intervals.
%In general, there is considerable variability in the appearance of the ITCZ and a band like (and thus very aggregated) structure of the moist region can be found during any time of the year, 
%Visual inspection of the CWV fields to determine what causes the reduced values of BLW suggests that, in spring and fall, the CWV field deviates from its stripe pattern in two different ways. 
%In Spring, they are frequently associated with high humidity values in the south-western part of the domain (see e.g., May 1st in Figure~\ref{fig:seasonal_cycle}b), while in fall the moist margin is  intermittently disturbed towards the North (see e.g., Sep 9th Figure~\ref{fig:seasonal_cycle}b). 
%Please note that, during episodes of low BLW$_\mathrm{S}$ the day-to-day variability is large, frequently exposing a band like structure (see e.g., June 20th), and corresponding high values of BLW.
In spring, the BLW$_\mathrm{S}$ reaches lower values because the CWV field frequently develops a second band of high humidity values located in the south-western part of the domain (see e.g., May 1st in Figure~\ref{fig:seasonal_cycle}b) whereas in summer it is more common to have a single band (see e.g., June 20th). The lower values of BLW$_\mathrm{S}$ in the fall are associated with the moist margin being regularly disturbed towards the North (see e.g., Sep 9th).

During boreal spring, the low values of BLW$_\mathrm{S}$ agree with previous studies which show that high variability in moisture and precipitation is common during this time of year \cite<see also Figure~1 in>{Chiang2002}. A possible explanation for the increase in variability in the CWV distribution is that the ITCZ, delimited by the moist margin, approaches its southernmost extent. As a consequence, land-ocean interactions with the South American Continent might impact the shape of the CWV distribution. An alternative explanation might be that the weaker SST gradient during this time of year more weakly constrains the location of the ITCZ. 
Low values of BLW$_\mathrm{S}$ during boreal fall may come from the moist margin's distortion by African Easterly Waves and Hurricanes (peak season in mid-September, see e.g. \citeA{Landsea1993}). In particular, the CWV fields with low values of BLW$_\mathrm{S}$ in fall often show moist filaments extending towards the north, i.e. into the previously dry regions, and dry filaments extending towards the south, i.e. into the previously moist region in a whirl-like pattern. 

\section{Conclusion\label{sec:Conclusion}}
%Text here ===>>>
In conclusion, the length of the moist tropical margin (MML) can quantify convective aggregation in both idealized simulations and realistic data of the tropical atmosphere: The shorter the MML, the more aggregated the atmosphere. We have shown that:
\begin{enumerate}
    \item The moist margin is well-identified using the anti-mode of the CWV distribution and easy to visualize (Figure~\ref{fig:Fig1}).
    \item The MML is linked to the evolution of the entire CWV distribution via the system's Gibbs free energy (Figure~\ref{fig:Fig2}).
    \item This allows us to formulate a robust index (BLW) to quantify convective aggregation across datasets and geometries (Figure~\ref{fig:BLW}).
    \item Unlike traditional aggregation metrics, BLW can track both the coalescence and coarsening of moist regions across a wide range of climates despite only requiring the choice of a fixed CWV percentile (Figure~\ref{fig:idealized}).
    \item BLW quantifies aggregation in realistic datasets, as shown by its ability to quantify the seasonal cycle of the Atlantic ITCZ's organization (Figure~\ref{fig:seasonal_cycle}).
\end{enumerate}

Compared to other convective organization indices such as SCAI \cite{tobin2012observational}, COP \cite{white2018quantifying}, or Iorg \cite{tompkins2017organization}, BLW is easy to interpret as the moist margin can be superimposed on the moisture field. Furthermore, BLW only relies on two transparent choices: (1) the fixed percentile defining the moist margin; and (2) the reference underlying shape of the moisture field. Both choices have default values facilitating the calculation of BLW: (1) the percentile can be chosen as the anti-mode of the CWV PDF if it is bimodal or as the maximum of the CWV PDF if it is unimodal; and (2) the reference underlying shape can be chosen as a circle for large square domains or as a band for elongated domain with anisotropic surface conditions. Compared to other shape-based organization indices \cite<e.g.,>[]{pscheidt2019organized,maceachren1985compactness}, BLW is strongly rooted in theory as we have shown in Section \ref{subsec:Contour_length} that the MML is closely related to the empirically-determined Gibbs free energy of the system. This link, analogous to surface tension, allows to relate BLW to the physical processes driving the evolution of the moisture field, hence opening the door to simple theories for the bimodal distribution of tropical water vapor \cite<e.g.,>[]{masunaga2020mechanism}. Note that we have focused on CWV rather than MSE for consistency with the moist margin's literature: As the CWV budget is dominated by advection, precipitation and evaporation, it makes the potential's shape hard to predict a priori, possibly reducing it to a diagnostic tool. Using MSE allows to bypass this limitation by decomposing the potential into diabatic components that can be individually interpreted, as we show in \ref{app:Decomposition}.

For conciseness, we have made several limiting choices throughout the manuscript. First, we have limited ourselves to the COSMO model for idealized RCE simulations because it leverages GPUs to run large ensembles on big domains at high resolution (see \ref{app:RCE_simulations}). While we have anecdotal evidence that the MML can successfully quantify convective organization in RCE simulations using different models \cite<e.g., models from>{Wing_2018}, BLW should be compared to other indices across models to test its broad applicability. Here, we have purposely avoided tuning BLW for inter-comparison purposes to keep a strong link with theory. Second, we have quantified organization in idealized simulations and reanalysis of different geometries to show the versatility of the moist margin's approach. However, this limits our ability to meaningfully compare idealized simulations to observations, as e.g. idealized simulations with weak rotation and meridional surface temperature gradients would be better analogues to the Atlantic ITCZ \cite<e.g.>{Muller2020}. 

Overall, our work generalizes the coarsening theory of \citeA{windmiller2019universality} to develop a mixed Eulerian-Lagrangian framework that simplifies (dis)aggregation processes to a competition between two opposing tendencies: (1) Aggregating tendencies that typically show a double-well potential structure and effectively act as a surface tension for the moist margin, and (2) Disaggregating tendencies that typically show a single-well potential structure and stretch or break the moist margin. Exciting opportunities lie ahead, such as evaluating this effective surface tension as an emergent constraint for the behavior of convective parametrizations, and using BLW to study the effect of aggregation on precipitation and climate sensitivity in atmospheric observations \cite<e.g.,>[]{popp2019impact}.

\appendix

\section{Three-dimensional RCE Simulations\label{app:RCE_simulations}}

%\subsubsection{Model Description}

The simulations were performed with the non-hydrostatic limited-area weather and climate model COSMO \cite<Consortium for Small-scale Modeling model, v5.0,>[]{Steppler_2003}.  In this model the thermo-hydrodynamical Euler equations are discretized on a structured longitude-latitude-height mesh using finite difference methods. The discretization schemes include a fifth-order upwind scheme for horizontal advection \cite{Baldauf_2011} and a split-explicit three-stage second-order Runge--Kutta time-stepping scheme for the forward integration in time \cite{Wicker_Skamarock_2002}. To damp gravity waves at the upper boundary, an implicit Rayleigh damping term on the vertical velocity is added at the end of each acoustic time step \cite{Klemp_2008}. 

Subgrid-scale processes include an interactive radiative transfer scheme based on the $\delta$-two-stream approach \cite{Ritter_1992} and a single-moment bulk cloud-microphysics scheme with five hydrometeor species \cite<cloud water, cloud ice, rain, snow, and graupel,>[]{Reinhardt_2006}. In the planetary boundary layer and for surface transfer a turbulent-kinetic-energy-based parameterization is used  \cite{Mellor_Yamada_1982, Raschendorfer_2001}, and parametrizations for convection have been switched off. 

The presented set of simulations have become possible due to a new version of COSMO capable of exploiting accelerators based on Graphics Processing Units (GPUs), which possess properties beneficial for weather and climate codes \cite{Owens_2008, Fuhrer_2014}. Our version supports executing the entire time stepping on GPU accelerators and thus avoids expensive data movements between host CPU and GPU accelerators. In practice, these capabilities enable kilometer-resolution simulations on near-global computational domains over extended periods of time \cite{Fuhrer_2018, Leutwyler_2016, Leutwyler_2019}. Here we exploit the capabilities to obtain a set of cloud-resolving RCE simulations at nine different sea surface temperatures (SST).

For the GPU version of COSMO, extensive validation has been conducted for kilometer-scale configurations. Examples include a ten-year-long reanalysis-driven simulation over Europe \cite{Leutwyler_2017}, validation of  clouds \cite{Hentgen_2019} and surface winds \cite{Belusic_2018}. Note that in our version of the COSMO model the atmospheric moisture budget is not strictly conserved. In the equilibrium phase of the simulation (last 40 days) about 0.069 kg/m$^2$/day of water is lost, i.e. about 1.5\% of the latent heatflux. 

% \tominline{Todo: Quantify Conservation of mass, energy and momentum to avoid the reviewers' ire}

%
%\section{Here Is Appendix Title}
% will show
% A: Here Is Appendix Title
%
%\appendix
\section{Decomposition of the Potential into Diabatic Tendencies and Advection\label{app:Decomposition}}

\begin{figure}
 \centering
\noindent\includegraphics[width=\textwidth]{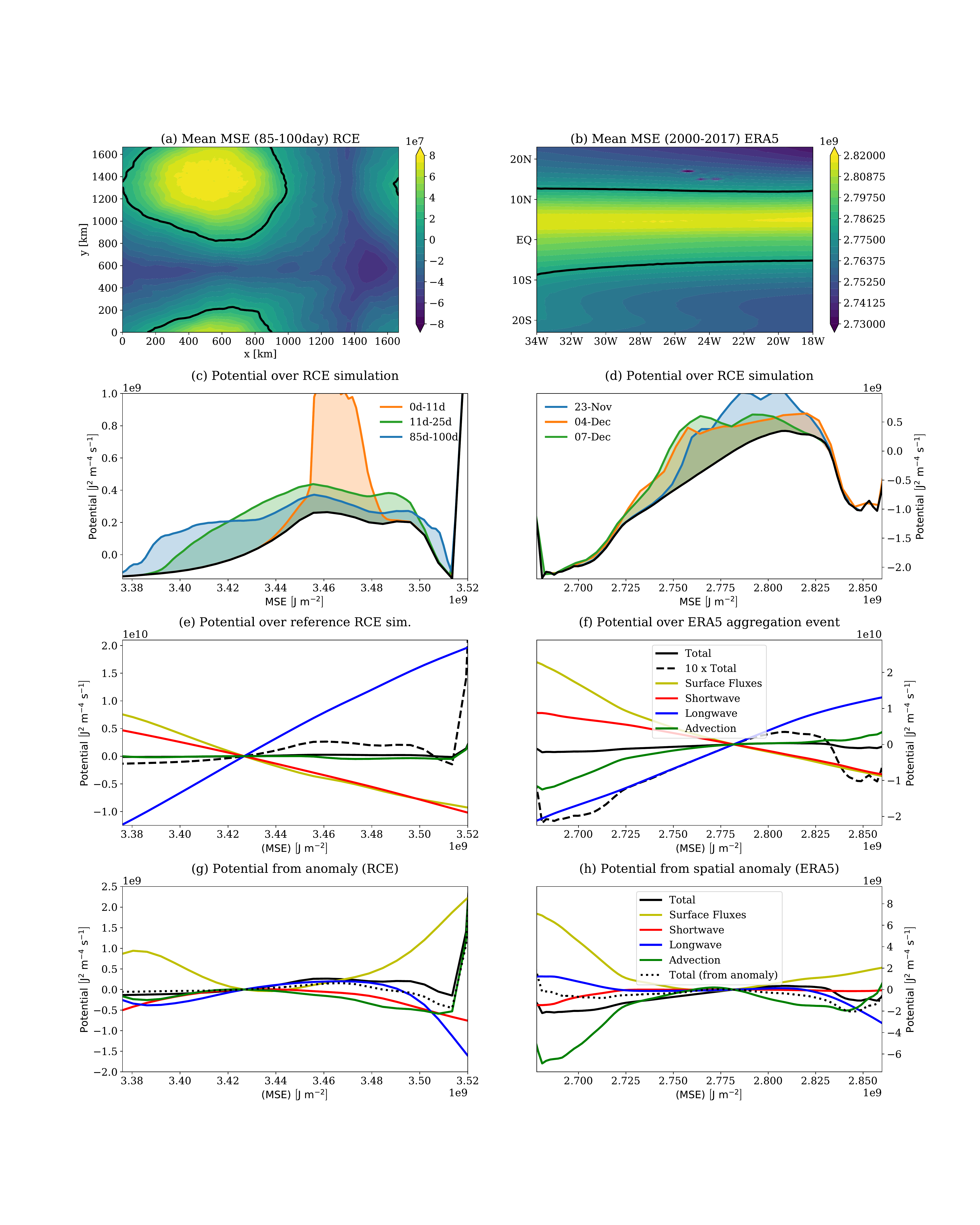}
 \caption{(a) Time-mean MSE field in RCE for day 85-100 with the 88th percentile marked in black. (b) Time-mean MSE field in ERA5 over 2000-2017 with the 67th percentile marked in black. (c) Potential for MSE calculated over the RCE simulation (black line), where we have added the scaled PDFs of MSE for different times using colored lines and shading. (d) Potential for MSE calculated between Nov 23rd and Dec 4th (black line), where we have added the scaled PDFs of MSE for different days using colored lines and shading. (e) Potential calculated for each term of equation \ref{eq:Decomposition} for RCE. (f) Potential calculated for each term of equation \ref{eq:Decomposition} for ERA5. (g) Potential calculated for each term of equation \ref{eq:Decomposition_anomaly} for RCE. (h) Potential calculated for each term of equation \ref{eq:Decomposition_anomaly} for ERA5. 
}
 \label{fig:Potential_decomposition}
\end{figure}

In this section, we choose column frozen moist static energy (MSE for short, in units $ \mathrm{W\ m^{-2}}$) as our order variable to decompose the potential into contributions from diabatic tendencies and MSE advection. Our starting point is the budget for column MSE, which we write in Eulerian form:

\begin{equation}
\frac{\partial\mathrm{MSE}}{\partial t}=\sum_{i=\mathrm{lw,sw,sef,adv}}\dot{\mathrm{MSE}}_{i}=-\sum_{i=\mathrm{lw,sw,sef,adv}}\frac{dV_{i}\left(\mathrm{MSE}\right)}{d\mathrm{MSE}},
    \label{eq:Decomposition}
\end{equation}

where we have decomposed the MSE tendency into its contribution from the net longwave atmospheric heating (lw), the net shortwave atmospheric heating (sw), the surface enthalpy fluxes (sef), and the horizontal flux of
MSE (adv, here calculated as a residual) through the grid cell's boundaries. We calculate a potential for each tendency using equation \ref{eq:calc_V} and depict the results in Figure~\ref{fig:Potential_decomposition}.

The first step is to re-define the moist margin to accommodate the change from CWV to MSE, which we do using the anti-mode of the MSE PDF for consistency with Section \ref{sec:Theory}. For the 300K idealized RCE simulation (Figure~\ref{fig:Potential_decomposition}a), we identify the anti-mode as the 88th percentile of the MSE PDF (Figure~\ref{fig:Potential_decomposition}c), which is the same as the percentile chosen for CWV (Figure~\ref{fig:Fig2}a) and underlines the strong constraint on weak temperature gradients in non-rotating RCE \cite<e.g.,>[]{sobel2000modeling}. In contrast, for the Tropical Atlantic reanalysis (Figure~\ref{fig:Potential_decomposition}b), the MSE PDF (Figure~\ref{fig:Potential_decomposition}d) is quite different from the CWV PDF (Figure~\ref{fig:Fig2}b) and only marginally bimodal on the most aggregated day (Dec 4th), whose anti-mode corresponds to the 67th percentile.

If the potential is directly decomposed using equation \ref{eq:potential}, the total potential is an order of magnitude smaller than its individual components. As a consequence, Figure~\ref{fig:Potential_decomposition}e and \ref{fig:Potential_decomposition}f highlight the well-known atmospheric energy balance to first order: For instance, the atmosphere cools to space via longwave radiation, resulting in a negative tendency, which translates to a positive slope that tries to reduce atmospheric MSE. This order-of-magnitude difference between the total potential and its individual contributions suggest that \textit{anomalies} from the atmospheric energy balance are responsible for the double-well shape of the total potential. This motivates investigating the budget for the spatial anomaly of MSE:

\begin{equation}
\frac{\partial\mathrm{MSE}^{\prime}}{\partial t}=\sum_{i=\mathrm{lw,sw,sef,adv}}\dot{\mathrm{MSE}}_{i}^{\prime}=-\sum_{i=\mathrm{lw,sw,sef,adv}}\frac{dV_{i}^{\prime}\left(\mathrm{MSE}\right)}{d\mathrm{MSE}},
    \label{eq:Decomposition_anomaly}
\end{equation}

where we have introduced the spatial anomaly $X^{\prime}\overset{\mathrm{def}}{=}X-\left\langle X\right\rangle _{{\cal D}} $ from the domain-average $\left\langle X\right\rangle _{{\cal D}} $ of a variable X. Note that equation \ref{eq:Decomposition_anomaly} is derived by taking the spatial anomaly of equation \ref{eq:Decomposition} without assuming that the domain-mean MSE is steady. This assumption does not hold in our case, leading to a difference between the total potential derived from the MSE anomaly (dotted black line in Figure~\ref{fig:Potential_decomposition}g and \ref{fig:Potential_decomposition}h) and the total potential derived from the total MSE (full black line). As both potentials are similar and exhibit a double-well structure, we proceed to analyze how individual components of the potential derived from MSE anomalies, which are now of the same order of magnitude as the total potential, sum up to a double-well structure. First, the ``walls'' of the potential for extreme MSE values come from the advection term, which smoothes out the largest MSE anomalies through overturning circulations \cite<e.g.,>[]{holloway2016sensitivity}. The ``moist well'' comes from the decrease in longwave cooling at high CWV values \cite<e.g.,>[]{beucler2016moisture}, which heats the moist regions and effectively moistens them because of weak buoyancy gradients. The ``dry well'' comes from both the increase in longwave cooling and the decrease in anomalous shortwave heating at low CWV values. Finally, surface enthalpy fluxes oppose the double-well tendency by damping near-surface enthalpy disequilibrium, favoring a single-well potential and hence a unimodal MSE distribution. Overall, both idealized RCE simulations and observations offer a consistent picture of the anomalous MSE tendencies responsible for the double-well potential structure. The main difference is a larger double-well potential for MSE advection in ERA5, which is expected as the Atlantic ITCZ is out of balance but requires further work as the MSE budget is not explicitly closed in ERA5 reanalysis. 

% \begin{equation}
% \partial_{t}\mathrm{MSE}=\sum_{i=\mathrm{lw,sw,sef,adv}}\dot{\mathrm{MSE}}_{i}=-\sum_{i=\mathrm{lw,sw,sef,adv}}V_{i}^{\prime}\left(\mathrm{MSE}\right),
%     \label{eq:Decomposition}
% \end{equation}

\section{Defining BLW Across Climates \label{app:percentile}}
\begin{figure}
 \centering
\noindent\includegraphics[width=\textwidth]{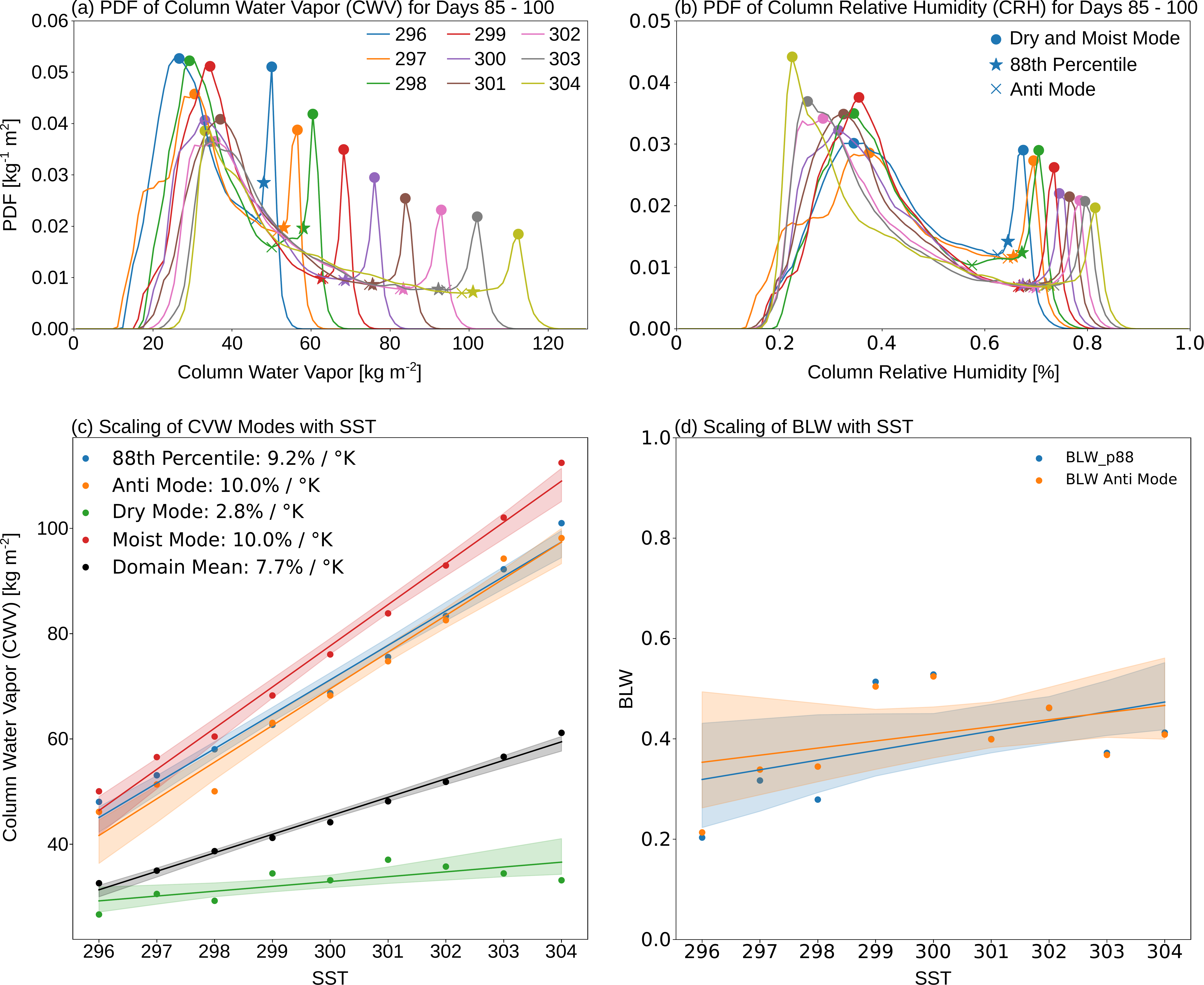}
 \caption{Scaling of CWV distribution modes with SST for the last 15 days of the RCE simulations. (a) CWV PDF at different SSTs (colored lines). The colored markers indicate the dry and moist local maxima (dots), the local minimum of the anti-mode (cross), and the 88th percentile (star). (b) Same as (a), but for column relative humidity.  (c) Scaling of the modes with SST. The solid lines represent the scaling using the mean slope (also indicated in the upper-left corner), and the shading the scaling using the 5th and 95th percentiles of the slope. The slope estimates were obtained by randomly choosing (with replacement) from the 9 SST samples.  (d) Same as (b), but for the final BLW$_C$ index. We use two different thresholds to define MML: The 88th percentile (blue dots) or the local minimum separating the two local maxima of the PDF (orange dots).}
 \label{fig:percentile_choice}
\end{figure}

% tgb - 6/2/2020 - I would argue that we can be a bit more general and see the definition of the anti-mode across SSTs as a special case of the definition of BLW across different climates
% We could make that point in the section's introduction
%DL-3-6-20: I did not fully understand what you meant, but made some changes.
Applying the BLW index entails choosing the idealized equilibrium shape of the moist region, and choosing a CWV contour representative of the potential hill separating the two potential wells.  The choice of a contour can be intricate as the different modes of the CWV distribution can change with climate (see Figure~\ref{fig:percentile_choice}) and because the anti-mode of the CWV distribution is not always well-defined. For example, the PDF's anti-mode is ill-defined when the CWV distribution exhibits multiple local minima (see e.g. green line in Figure~\ref{fig:percentile_choice}a). For ERA5, a similar problem arises, since the CWV distribution occasionally becomes uni-modal (Figure~\ref{fig:Fig1}), making an index based on the local minimum undefined.

To address this issue, we adopt a definition based on a temporally-fixed percentile of CWV, meaning that the domain is split into a moist and a dry region with temporally-fixed areas. Motivated by the study of \citeA{Jakob_2019}, the 88th percentile of CWV (wet-dry ratio of about 1:9) was chosen for the RCE simulation. For the Atlantic ITCZ, the 83rd percentile of CWV (ratio of about 1:6) was chosen, since it corresponds to the 48\,kg/m$^2$ contour proposed by \citeA{Mapes2018}. We discuss below the robustness of using the 88th percentile of CWV to represent the anti-mode for RCE.

The 88th percentile of CWV tracks the minimum of the CWV PDF reasonably well across a range of SSTs (Figures~\ref{fig:percentile_choice} and \ref{fig:Fig1}c). For the equilibrium phase (Day 85--100), the contour is akin to the minimum at an SST of 299--304\,K, while it resides on the hillside of the moist mode for colder SSTs (296 and 298\,K). While the match between the minimum and 88th percentile is not perfect, the two indices scale approximately at the same rate with SST (9.2\%\,K$^{-1}$ and 10\%\,K$^{-1}$). That rate is faster than that of the mean CWV (7.7\%K$^{-1}$) or the dry mode (2.8\%\,K$^{-1}$), and also matches that of the moist mode (10\%\,K$^{-1}$). Note, that the differential scaling of the two modes leads to a less sharp MML at warmer SSTs, since the valley separating the two maxima becomes wider.

The shape of the CWV distribution provides first indication of the organization in the RCE simulations. For instance, the distance between the marker of the 88th percentile (star) and the line of zero PDF (x-axis) is representative of the MML. The analysis presented in Figures~\ref{fig:percentile_choice} indicates that the MML is slightly longer at cold SSTs (296--298\,K), but remains almost at the same length for SSTs between 299 and 304\,K. In practice, the difference between the three cold simulations and the remaining set emerges from the number of clusters. In the colder simulations, we detect two clusters, while we find only one in the warmer simulations. It is thus only in the warm simulations that variations in BLW describe how the final cluster approaches a circular shape. Based on BLW, the 299 and the 300\,K simulations can be considered the most organized.

The geometrical perspective of the BLW index gives additional information compared to established aggregation indices. For instance, the popular IQR metric applied to column relative humidity would suggest that warmer simulations are more organized because their IQR is larger (Figure~\ref{fig:percentile_choice}b). This can be also be seen by looking at the CWV PDF (Figure~\ref{fig:percentile_choice}a), where the dry mode moves to the left and the moist mode moves to the right. In contrast, the geometric perspective of BLW suggests little to no change in aggregation with warming (Figure~\ref{fig:percentile_choice}d) as convection clusters in a more circular shape at SST=300K than at SST=304K despite the increase in column relative humidity variance at 304K.

\section{Potential and MML Timescales\label{app:Timescales}}

In this section, our goal is to estimate (1) the timescale to reach local equilibrium (fast formation of small-scale moist and dry regions); and (2) the timescale to reach global equilibrium in RCE (slow formation of a single moist cluster, Figure \ref{fig:Fig1}a), and to form the ``ITCZ stripe'' in ERA5 (Figure \ref{fig:Fig1}b). The timescale of local equilibrium ($\tau_{\mathrm{MML}}$) corresponds to fast time-variations of the MML defined in Section \ref{subsec:Contour_length}, while the timescale of global equilibrium ($\tau_{V}$) is governed by the empirical potential defined in Section \ref{sub:Potential}. In the absence of an exact theory, we diagnose both timescales from data and require that the timescales minimally depend on the spatio-temporal resolution of each data set and describe the non-monotonic oscillations of the observed MML.

\begin{figure}
 \centering
\noindent\includegraphics[width=\textwidth]{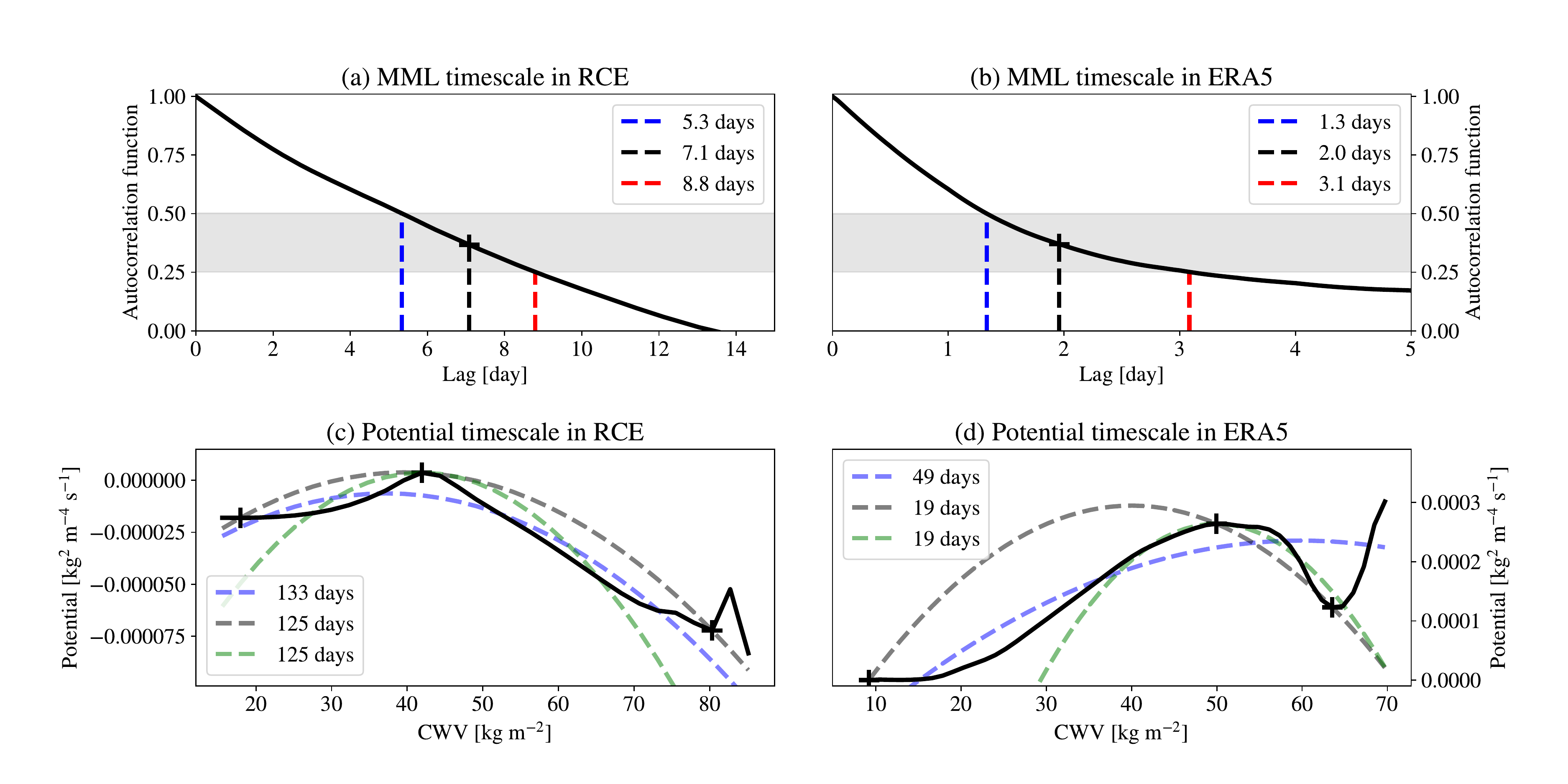}
 \caption{Three estimates of the MML timescale (vertical dashed lines) using the normalized autocorrelation function (black line) for RCE (panel a) and ERA5 (panel b). Three estimates of the potential timescale using equation \ref{eq:tau_V} and three polynomial approximations (dashed lines) for the potential (full black line) for RCE (panel c) and ERA5 (panel d).  \label{fig:app_timescales}}
\end{figure}

To estimate the MML timescale, we choose the e-folding timescale of the autocorrelation function, defined as the inverse Fourier transform of the MML's temporal power spectrum (e.g., Section 8.10 of \citeA{dunn2017measurement}). The autocorrelation function of the MML, depicted in Figure~\ref{fig:app_timescales}a for RCE and Figure~\ref{fig:app_timescales}b for ERA5, is normalized by its maximum to yield values between -1 and 1. We put uncertainty bounds on the e-folding timescale $\tau_{\mathrm{MML}}$ by additionally finding the shortest lag for which the normalized autocorrelation function is 0.5 and 0.25, leading to estimates of $\tau_{\mathrm{MML}}=\left(7.1\pm1.8\right)\mathrm{day}$ for RCE and   $\tau_{\mathrm{MML}}=\left(2.0\pm1.1\right)\mathrm{day}$ for ERA5. 

To estimate the potential timescale, we linearize the time-evolution equation \ref{eq:potential} about the CWV value $\mathrm{CWV_{max}}$ where the potential reaches its maximal value $\mathrm{V_{max}}$:

\begin{equation}
\frac{\partial\mathrm{CWV}}{\partial t}\approx-\left(\frac{d^{2}V}{d\mathrm{CWV^{2}}}\right)_{\mathrm{CWV_{max}}}\times\left(\mathrm{CWV}-\mathrm{CWV_{max}}\right).
\label{eq:Lin_Potential}
\end{equation}

Equation \ref{eq:Lin_Potential} describes a scenario where an initially uniform CWV field of intermediate CWV value $\mathrm{CWV_{max}}$, subjected to the bimodal potential $\mathrm{V}$ with local minima $V_{1}\overset{\mathrm{def}}{=}V\left(\mathrm{CWV}_{1}\right)$ and $V_{2}\overset{\mathrm{def}}{=}V\left(\mathrm{CWV}_{2}\right)$, evolves towards a self-aggregated state. The linear timescale dictating the evolution of CWV is given by the inverse of the potential's second derivative evaluated at $\mathrm{CWV_{max}}$:
\begin{equation}
\tau_{V}\overset{\mathrm{def}}{=}-\left(\frac{d^{2}V}{d\mathrm{CWV^{2}}}\right)_{\mathrm{CWV_{max}}}^{-1}.
\label{eq:tau_V}
\end{equation}

To calculate $\tau_{V}\ $from equation \ref{eq:tau_V}, we estimate the second derivative of $V$ using three different methods for robustness:
\begin{enumerate}
\item We fit a second-order polynomial (dashed blue lines in Figure~\ref{fig:app_timescales}) using all potential values between $\mathrm{CWV_{1}}$ and $\mathrm{CWV_{2}}$ (the left and right black crosses in Figure Figure~\ref{fig:app_timescales}) and estimate $\tau_{V}\ $ from its leading coefficient. 
\item We fit a second-order polynomial (dashed gray lines in Figure~\ref{fig:app_timescales}) using the three local extrema of the potential function (black crosses in Figure~\ref{fig:app_timescales}) and estimate $\tau_{V}\ $ from its leading coefficient.
\item We approximate the local curvature of the potential V (dashed green lines in Figure~\ref{fig:app_timescales}) using a second-order Taylor-Series expansion at $\mathrm{CWV_{1}}$, $\mathrm{CWV_{max}}$, and $\mathrm{CWV_{2}}$, from which we estimate $\tau_{V}\ $.
\end{enumerate}

Reassuringly, the mathematically-equivalent second and third methods give the same results to good precision, while the first method gives an upper bound for $\tau_{V}\ $, leading to estimates of $\tau_{V}=\left(125-133\right)\mathrm{day}$ for RCE and   $\tau_{V}=\left(19-49\right)\mathrm{day}$ for ERA5.

We can now estimate the ratio of the short timescale $\tau_{\mathrm{MML}}\ $ to the long timescale $\tau_{V}\ $: $\tau_{\mathrm{MML}}/\tau_{V}\approx\left(4-7\right)\%$ for RCE and $\tau_{\mathrm{MML}}/\tau_{V}\approx\left(3-16\right)\%$ for ERA5. This confirms that the timescales of local and global equilibria are separated by one to two orders of magnitude for both data sets, despite the real tropics evolving approximately five times faster than RCE.

\acknowledgments
We thank the 2nd ICTP Summer School on Theory, Mechanisms and Hierarchical Modelling of Climate Dynamics and the Max-Planck Institute for Meteorology for incubating the research project that led to this manuscript. We additionally thank Cathy Hohenegger, Bjorn Stevens, George Craig, Nicholas Lutsko, Jiawei Bao, Thibaut Dauhut, Hauke Schulz, Da Yang, and two anonymous reviewers for advice that improved the present manuscript. We acknowledge NSF grants OAC-1835769,  OAC-1835863 and AGS-1734164 for Tom Beucler's funding, the Swiss National Science Foundation under project No.  P2EZP2\_178503 for David Leutwyler's funding, and the Hans Ertel Centre for Weather Research for Julia Windmiller's funding. 
%, a German research network consisting of universities, research institutions and the Deutscher Wetterdienst (German Meteorological Service) and which is funded by the Bundesministerium für Verkehr und digitale Infrastruktur (Federal Ministry of Transport and Digital Infrastructure).

The Github repository for our project can be found at \url{https://github.com/tbeucler/2019_WMI}. The RCE simulations were conducted at the Swiss National Supercomputing Centre CSCS. For the particular version of COSMO, we would like to thank the Federal Office of Meteorology and Climatology MeteoSwiss, the Centre for Climate Systems Modeling (C2SM) and ETH Zurich. In particular we would like to acknowledge Linda Schlemmer for providing us with the initial set of code modifications for RCE. COSMO may be used for operational and for research applications by the members of the COSMO consortium. Moreover, within a license agreement, the COSMO model may be used for operational and research applications by other national (hydro-)meteorological services, universities, and research institutes. The particular version of the COSMO model used in this study is based on the official version 5.0 with many additions to enable GPU capability and available under a license (\url{http://www.cosmo-model.org/content/consortium/licencing.htm}). The ERA5 reanalysis dataset, provided by the Copernicus Climate Change Service (C3S), was directly downloaded from the Copernicus Climate Change Service Climate Data Store (CDS, \url{https://cds.climate.copernicus.eu/cdsapp#!/home}) in October 2018 on the Engaging computing cluster provided by MIT before being copied over to CSCS.

%% ------------------------------------------------------------------------ %%
%% References and Citations

%%%%%%%%%%%%%%%%%%%%%%%%%%%%%%%%%%%%%%%%%%%%%%%
%
% \bibliography{<name of your .bib file>} don't specify the file extension
%
% don't specify bibliographystyle
%%%%%%%%%%%%%%%%%%%%%%%%%%%%%%%%%%%%%%%%%%%%%%%

\bibliography{main.bbl}

%Reference citation instructions and examples:
%
% Please use ONLY \cite and \citeA for reference citations.
% \cite for parenthetical references
% ...as shown in recent studies (Simpson et al., 2019)
% \citeA for in-text citations
% ...Simpson et al. (2019) have shown...
%
%
%...as shown by \citeA{jskilby}.
%...as shown by \citeA{lewin76}, \citeA{carson86}, \citeA{bartoldy02}, and \citeA{rinaldi03}.
%...has been shown \cite{jskilbye}.
%...has been shown \cite{lewin76,carson86,bartoldy02,rinaldi03}.
%... \cite <i.e.>[]{lewin76,carson86,bartoldy02,rinaldi03}.
%...has been shown by \cite <e.g.,>[and others]{lewin76}.
%
% apacite uses < > for prenotes and [ ] for postnotes
% DO NOT use other cite commands (e.g., \citeA, \citep, \citeyear, \nocite, \citealp, etc.).
%

\end{document}